\documentclass[12pt]{iopart}
\usepackage{iopams}

\expandafter\let\csname equation*\endcsname\relax
\expandafter\let\csname endequation*\endcsname\relax

\usepackage{amsmath}
\usepackage{amssymb}
\usepackage{amsthm}

\usepackage[a4paper,left=2.2cm,right=2.2cm,top=3cm,bottom=3cm,footnotesep=1.5cm]{geometry}

\usepackage{kantlipsum}

\makeatletter
\def\footnoterule{\kern-3mm\kern-3\p@
  \hrule \@width 2in \kern 2.6\p@
  \kern 3mm}
\makeatother

\usepackage[english]{babel}

\renewcommand{\figurename}{FIG.}
\addto\captionsenglish{\renewcommand{\figurename}{FIG.}}

\renewcommand{\tablename}{TABLE}
\addto\captionsenglish{\renewcommand{\tablename}{TABLE}}

\usepackage{subfigure}

\usepackage{comment}
\usepackage{array}
\usepackage{multirow}
\usepackage{relsize}
\usepackage{dsfont}

\usepackage[colorlinks]{hyperref}
\hypersetup{
	colorlinks = true,
	linkcolor = red,
	citecolor = {green},
	urlcolor = {blue}
}

\newcounter{Sequ}
\newenvironment{SEquation}
    {\stepcounter{Sequ}%
     \addtocounter{equation}{-1}%
     \renewcommand\theequation{\Alph{section}.\arabic{Sequ}}\equation}
    {\endequation}

\usepackage{graphicx}
\usepackage{caption}
\usepackage{lipsum}
\usepackage[normalem]{ulem}
\usepackage{dcolumn}

\usepackage{bm}

\usepackage[square,numbers,sort&compress]{natbib}

\usepackage{fancyhdr}
\begin{document}

\title{Impurity Dephasing in a Bose-Hubbard Model}

\author{Fabio Caleffi$^1$}
\ead{fabio.caleffi@sissa.it}
\vspace{-10pt}
\author{Massimo Capone$^{1, 2}$}
\vspace{-10pt}
\author{Inés de Vega$^{3,4}$}
\vspace{-10pt}
\author{Alessio Recati$^{5, 6}$}
\ead{alessio.recati@unitn.it}

\vspace{10pt}

\address{$^1$International School for Advanced Studies (SISSA), Via Bonomea 265, I-34136 Trieste, Italy}
\address{$^2$CNR-IOM Democritos, Via Bonomea 265, I-34136 Trieste, Italy}
\address{$^3$Department  of  Physics  and  Arnold  Sommerfeld  Center  for  Theoretical  Physics, Ludwig-Maximilians-Universit\"at M\"unchen,  Theresienstr. 37,  80333 Munich, Germany}
\address{$^4$IQM Germany GmbH, Nymphenburgerstr. 86 80636 Munich, Germany}
\address{$^5$INO-CNR  BEC  Center  and  Dipartimento di  Fisica,  Universit\`a di Trento,  38123  Povo, Italy}
\address{$^6$Trento Institute for Fundamental Physics and Applications, INFN, 38123, Trento, Italy}

\vspace{10pt}

\begin{indented}
\item[] \today
\end{indented}

\begin{abstract}
We study the dynamics of a two-level impurity embedded in a two-dimensional Bose-Hubbard model at zero temperature from an open quantum system perspective.
Results for the decoherence across the whole phase diagram are presented, with a focus on the critical region close to the transition between superfluid and Mott insulator.
In particular we show how the decoherence and the deviation from a Markovian behavior are sensitive to whether the transition is crossed at commensurate or incommensurate densities.
The role of the spectrum of the Bose-Hubbard environment and its non-Gaussian statistics, beyond the standard independent boson model, is highlighted.

Our analysis resorts on a recently developed method [Phys. Rev. Research 2, 033276 (2020)] -- closely related to slave boson approaches -- that enables us to capture the correlations across the whole phase diagram. This semi-analytical method provides us with a deep insight into the physics of the spin decoherence in the superfluid and Mott phases as well as close to the phase transitions.
\end{abstract}

\section{Introduction}
Understanding the dynamics of an open quantum system, i.e., a quantum system coupled to its environment, is relevant in a variety of domains including condensed matter physics, quantum computing, quantum optics and ultracold gases \cite{breuer,devega2015c,rivas2011a,rivas2014}. When the open system and its environment are weakly coupled, it is often a good approximation to describe the latter as a set of harmonic oscillators linearly coupled to the system. This class of problems is well described by the so-called Caldeira-Leggett model, when the open system is described in terms of continuous variables, or by the spin-boson model, when it is a discrete system. In any of these models, the influence of the environment on the system depends only on a single-particle spectral density, and this strongly simplifies the description of the system. The past few decades have seen the development of a large variety of methods to describe the open system dynamics in this context, including path integrals \cite{strunz1996,makri1998}, stochastic Schr\"odinger equations \cite{diosi1998,alonso2005}, hierarchical systems of equations \cite{tanimura20151,changyu2018a} or, when computing the full dynamics of both the system and its environment, chain mapping representations \cite{prior2010,devega2015} or quantum Monte Carlo techniques \cite{muehlbacher2006,pollet2012}.

However, when the environment is strongly correlated or non-harmonic, the above picture may no longer be accurate and more involved approaches are required to account for the resulting non-Gaussian environment statistics. The state-of-the-art methods to numerically study these systems are based on matrix-product states \cite{scholl2005, scholl2011, cosco2018, bramberger2020}; nevertheless, due to the rapid entanglement growth, these methods become highly inefficient beyond one-dimensional cases or when approaching to a critical regime.

The recent advances in locally manipulating ultra-cold gases in optical lattices has made such a platform ideal for the study of impurities coupled to a non-trivial bath
\cite{streif2016,kantian2015,mitchison2016,cosco2018,giorgi2019,vardhan2017} either per se or as quantum simulators of toy models for less clean systems. In this paper, we analyze the pure dephasing dynamics of a two-level impurity whose environment is represented by a single-band Bose-Hubbard (BH) model. This problem has been recently analyzed for a one-dimensional BH environment away from its critical transition \cite{cosco2018}. Here we take a leap forward by considering a 2D BH model and characterizing the impurity dynamics along the whole phase diagram, focusing on the critical regions. Our goal can be reached thanks to the use of a Gutzwiller technique that we recently developed \cite{caleffi2019quantum}. The method  allows us to include the relevant correlations of the bath -- in particular the ones responsible for non-Gaussian effects -- without being computationally demanding.

One of the main findings of our study is the strong dependence of the dephasing dynamics on the universality class of the Mott insulator-superfluid transition of the BH environment.
In particular, we show that: (1) when the quantum phase transition is due to particle number change, also known as commensurate-incommensurate transition, the impurity dynamics is perfectly Markovian, being the environment dynamics dominated by single particle processes, despite the strong interactions; (2) on the other hand, when the transition occurs at fixed (integer) density, the spectrum of the bath contains multiple low-energy collective modes. Their presence leads to a non-Markovian dephasing dynamics, strongly affected by two-particle processes in the environment, which make the standard Gaussian statistics fail. Most importantly -- in close analogy with the findings of a related work on one-dimensional quantum spin baths \cite{fazio2006} -- we find that both the short and long-time behaviour of the dephasing dynamics are precise detectors of the type of universality class of the transition.

The paper is organized as follows. Section~\ref{sec_model&theory} is devoted to introducing the pure dephasing model, the quantum Gutzwiller approach used to access the relevant Bose-Hubbard correlations and the so-called BLP non-Markovianity measure of dephasing processes, which is taken as a reference for our analysis. In Section~\ref{sec_results}, we present our predictions for the dephasing dynamics across the phase diagram of the Bose-Hubbard environment, focusing on the intrinsic non-Markovian effects due to the lattice setting and the consequences of the spectral properties of the bath. Specifically, the role of the superfluid-Mott insulator transition is highlighted. We conclude in Section~\ref{sec_summary} including an outlook on future studies of experimental interest.

\section{Model and theory}\label{sec_model&theory}
\subsection{Quantum impurity in a Bose-Hubbard bath}\label{subsec_model}
We consider a two-level impurity coupled to a two-dimensional single-band Bose-Hubbard (BH) model \cite{fisher, sachdev} with Hamiltonian $\hat{H}_{BH}$, hereafter referred to as the bath. The total Hamiltonian of the system can be written as $\hat{H} = \hat{H}_{BH} + \hat{H}_{imp} + \hat{H}_c$ with 
\begin{equation}\label{model}
\begin{gathered}
\hat{H}_{BH} = -J \sum_{\langle \mathbf{r}, \mathbf{s} \rangle} \left( \hat{a}^{\dagger}_{\mathbf{r}} \, \hat{a}_{\mathbf{s}} + \text{h.c.} \right) + \frac{U}{2} \sum_{\mathbf{r}} \hat{n}_{\mathbf{r}} \left( \hat{n}_{\mathbf{r}} - 1 \right) - \mu \sum_{\mathbf{r}} \hat{n}_{\mathbf{r}} \, , \\
\hat{H}_{imp} =  \frac{\hbar \, \omega_0}{2} \left( 1 + \hat{\sigma}_z \right) \, , \\
\hat{H}_c = g \, \hat{\sigma}_z \, \hat{n}_{\mathbf{0}} \, ,
\end{gathered}
\end{equation}
where the operators $\hat{a}_{\mathbf{r}} \left( \hat{a}^{\dagger}_{\mathbf{r}} \right)$ annihilate (create) a boson on the lattice site $\mathbf{r}$, $J$ is the hopping energy, $U$ the on-site bath interaction and $\mu$ the chemical potential, while $\langle \mathbf{r}, \mathbf{s} \rangle$ labels all pairs of nearest-neighboring sites. The impurity is assumed to be located at site $\mathbf{0}$ at the centre of a thermodynamically large BH lattice; it is governed by the Hamiltonian $\hat{H}_{imp}$ with a resonant frequency $\omega_0$ and is coupled to the bath density $\hat{n}_{\mathbf{0}}$ via a local interaction $\hat{H}_c$ with strength $g$.

We assume that initially the system's state is separable $\rho(t = 0) = \rho_{BH}^0 \otimes \rho_{imp}^0$, where $\rho_{BH}^0$ is the zero-temperature ground state of the BH Hamiltonian $\hat{H}_{BH}$ and $\rho_{imp}^0$ is the initial state of the impurity.
As usual in the study of open quantum systems, we assume that the bath and the impurity are weakly coupled so that the bath's state is not too altered with respect to $\rho_{BH}^0$. Under such approximation, it is well-known that the impurity dynamics is fully characterized by the time correlation function of the environment coupling operator, $\hat{n}_{\mathbf{0}}$. We estimate the latter by using a recently developed quantum Gutzwiller (QGW) approach \cite{caleffi2019quantum}, that has been proven to be very accurate to describe the quantum correlations of the BH model across the whole phase diagram. We refer the reader to the original paper \cite{caleffi2019quantum} and to \ref{app_qga} where we briefly review the method.

\subsection{The Quantum Gutzwiller method}\label{subsec_qga}


Within the QGW, the BH environment -- aside from a constant energy term -- can be recast as the quadratic Hamiltonian 
\begin{equation}\label{H_QGA_text}
\hat{H}^{\left( 2 \right)}_{QGW} = \hbar\sum_{\alpha} \sum_{\mathbf{k}} \omega_{\alpha, \mathbf{k}} \, \hat{b}^{\dagger}_{\alpha, \mathbf{k}} \hat{b}_{\alpha, \mathbf{k}} \, ,
\end{equation}
where the operator $\hat{b}_{\alpha, \mathbf{k}}$ ($\hat{b}_{\alpha, \mathbf{k}}^\dagger$) annihilates (creates) an excitation in the branch $\alpha$ with momentum $\mathbf{k}$, whose energy is $\hbar \, \omega_{\alpha,\mathbf{k}}$. 
The quadratic nature of the bath Hamiltonian allows us to easily estimate its quantum correlations.

\begin{figure}[!t]
    \centering
	\includegraphics[width=0.8\linewidth]{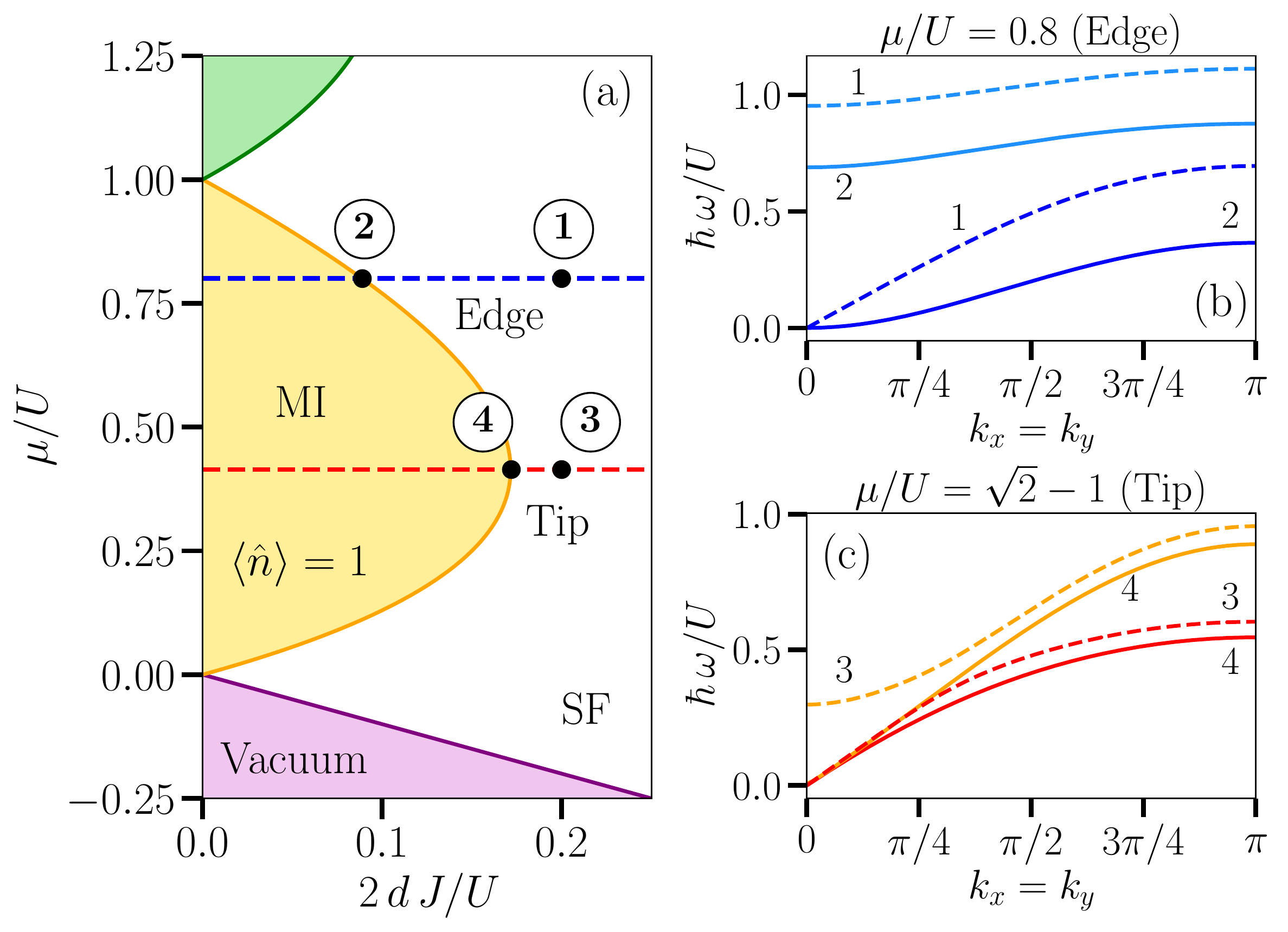}
    \caption{Panel (a): mean-field phase diagram of the BH model around the Mott lobe with integer filling $\langle \hat{n} \rangle = 1$. The blue (red) dashed line marks the path crossing the incommensurate (commensurate) MI-SF transition point considered in this work. The two panels (b) and (c) at the right hand side depict the energy dispersion at the points (1)-(4) represented in the phase diagram (a).
    Panel (b), represents points (1) and (2) near and within the critical point at the edge transition, in dashed and solid lines respectively. For these two points, the Goldstone and Higgs modes are represented in blue and light-blue lines respectively.
    Panel (c) represents points (3) and (4) near and at the critical point at the tip transition, again in dashed and solid lines respectively. Goldstone and Higgs modes are now represented in red and orange lines respectively.}
    \label{phase_diagram}
\end{figure}

Before proceeding, we briefly review the structure of the BH excitation spectrum $\omega_{\alpha, \mathbf{k}}$ along the phase diagram, since its knowledge gives important insights in the dephasing dynamics of the spin impurity, as we show in Section~\ref{sec_results}. The spectrum is well-known and can be obtained also from linear-response theory applied to the time-dependent Gutzwiller approximation \cite{krutitsky_navez, caleffi2019quantum}. 
For convenience, in \autoref{phase_diagram} a summary of the phase diagram and of the excitation spectra in different regimes is shown.
The most relevant feature of the BH model is the existence of a quantum phase transition between a Mott insulator (MI) -- which favours localized particles and occurs at integer fillings for $U/J$ larger than a critical value -- and a superfluid (SF) delocalized phase with broken U(1) symmetry. The quantum criticality is characterized by two different universality classes \cite{fisher, sachdev}, depending on whether the transition point is crossed by tuning the density to a commensurate (i.e., integer) lattice filling -- the so-called commensurate-incommensurate (CI) transition [at the edge of the Mott lobe: see point 2 on the blue dashed line in \autoref{phase_diagram}(a)]  -- or it is crossed at a fixed commensurate filling [at the tip of the Mott lobe: see point 4 on the red dashed line in \autoref{phase_diagram}(a)] -- crossing the so-called $\text{O}{\left( 2 \right)}$ transition. 

In the MI incompressible phase, the two lowest excitation branches are the gapped particle and hole excitations (not shown in \autoref{phase_diagram}). As the SF phase is approached along a CI transition line one of the excitations becomes gapless and transforms into the superfluid gapless Goldstone mode. 
The low momentum dispersion relation of the Goldstone mode becomes quadratic
at the transition point, while is linear in the SF phase (collisionless sound mode) [\autoref{phase_diagram}(b)]. Therefore, at the CI critical point the BH system, although strongly interacting, behaves as a free Bose gas of quasi-particles.

Instead, at the fixed-density $\text{O}{\left( 2 \right)}$ critical point, both the lowest-energy modes are gapless [\autoref{phase_diagram}(c)], and, in sharp contrast with the CI critical region, have a linear dispersion relation. In the SF phase only one linear gapless mode is present with finite sound velocity [\autoref{phase_diagram}(c)]. The other gapped excitation is often referred to as the Higgs mode and it is related to the amplitude fluctuations of the order parameter \cite{huber, BEC}. 

The QGW approach provides a recipe to express operators and observables of the BH bath in terms of the excitations operators $\hat{b}_{\alpha, \mathbf{k}}$ (see \cite{caleffi2019quantum} and \ref{app_qga}).
In particular, the impurity dynamics due to the weak coupling with the bath as described by Eq.~\eqref{H_QGA_text} is fully characterised by the time dependent density correlation function at the impurity position. The expression for the density operator $\hat{n}_\mathbf{0}$ can be written within the QGW approach as
\begin{equation}\label{density_expansion}
\hat{n}_{\mathbf{0}} \approx n_0 + \delta_1 \hat{n}_{\mathbf{0}} + \delta_2 \hat{n}_{\mathbf{0}} \, ,
\end{equation}
where $n_0$ is the mean-field density and we separate the single quasi-particle contribution  
\begin{equation}\label{BH_density_1}
\delta_1 \hat{n}_{\mathbf{0}}{\left( t \right)} = \frac{1}{\sqrt{V}} \sum_{\alpha} \sum_{\mathbf{k}} N_{\alpha, \mathbf{k}} \left( e^{-i \, \omega_{\alpha, \mathbf{k}} \, t} \, \hat{b}_{\alpha, \mathbf{k}} + e^{i \, \omega_{\alpha, \mathbf{k}} \, t} \, \hat{b}^{\dagger}_{\alpha, \mathbf{k}} \right) \, ,
\end{equation}
from the two-particle contribution  
\begin{equation}\label{BH_density_2}
\begin{aligned}
\delta_2 \hat{n}_{\mathbf{0}}{\left( t \right)} = \frac{1}{V} \sum_{\alpha, \beta} \sum_{\mathbf{k}, \mathbf{p}} \bigg[ &W_{\alpha \mathbf{k}, \beta \mathbf{p}} \, e^{i \left( \omega_{\alpha, \mathbf{k}} + \omega_{\beta, \mathbf{p}} \right) t} \, \hat{b}^{\dagger}_{\alpha, \mathbf{k}} \, \hat{b}^{\dagger}_{\beta, \mathbf{p}} + W_{\beta \mathbf{p}, \alpha \mathbf{k}} \, e^{-i \left( \omega_{\alpha, \mathbf{k}} + \omega_{\beta, \mathbf{p}} \right) t} \, \hat{b}_{\alpha, \mathbf{k}} \, \hat{b}_{\beta, \mathbf{p}} \\
&+ U_{\alpha \mathbf{k}, \beta \mathbf{p}} \, e^{i \left( \omega_{\alpha, \mathbf{k}} - \omega_{\beta, \mathbf{p}} \right) t} \, \hat{b}^{\dagger}_{\alpha, \mathbf{k}} \, \hat{b}_{\beta, \mathbf{p}} + V_{\alpha \mathbf{k}, \beta \mathbf{p}} \, e^{-i \left( \omega_{\alpha, \mathbf{k}} - \omega_{\beta, \mathbf{p}} \right) t} \, \hat{b}_{\alpha, \mathbf{k}} \, \hat{b}^{\dagger}_{\beta, \mathbf{p}} \bigg] \, ,
\end{aligned}
\end{equation}
where $V$ is the lattice volume. The coefficients $N_{\alpha, \mathbf{k}}$ and $W_{\alpha \mathbf{k}, \beta \mathbf{p}}$, $U_{\alpha \mathbf{k}, \beta \mathbf{p}}, V_{\alpha \mathbf{k}, \beta \mathbf{p}}$ are given explicitly in \ref{app_qga} and correspond to the spectral decomposition of the single and two-particle structure factors of density correlations in the Bose-Hubbard system.

It is worth noticing that the inclusion of two-particle processes due to $\delta_2 \hat{n}$ into the bath description generalizes the independent boson model, where the impurity polarization $\hat{\sigma}_z$ couples only to linear contributions of the form \eqref{BH_density_1} (see, e.g., \cite{MahanBook}). Indeed, we underline that the two-particle contributions dominate the density correlation functions in the MI phase and close to the MI-SF transition \cite{caleffi2019quantum}. In the following we show that this is the case also for the impurity dephasing, but not at the CI transition point. 

Let us stress that, compared to other approaches like strong-interaction perturbative methods \cite{bruder} and the standard Bogoliubov approximation \cite{rey}, the QGW approach provides a unified description from the deep MI state to the weakly-interacting superfluids. Moreover, it yields an insight into the spectral composition of quantum expectation values.

\subsection{Non-Markovianity measure of pure dephasing}\label{subsec_markovianity}
Having reduced the BH environment to the effective quadratic model \eqref{H_QGA_text}, the theoretical investigation of pure dephasing dynamics becomes tractable in the limit in which the presence of the impurity does not perturb significantly the behaviour of the environment, i.e. when the bath-impurity coupling $g$ is small compared to all the other energy scales of the problem. For the purpose of this study, we choose to work in such weak coupling limit. Using the time-convolutionless projection operator technique up to second order in the coupling constant $g$ \cite{breuer}, the evolution of the density matrix of the impurity is proved to obey a time-local master equation \cite{doll2008}
\begin{equation}\label{master_equation}
\partial_t \, \rho_{imp} = -i \frac{\tilde{\omega}_0}{2} \left[ \hat{\sigma}_z, \rho_{imp} \right] + \frac{g^2}{2 \, \hbar^2} \gamma{\left( t \right)} \left( \hat{\sigma}_z \, \rho_{imp} \, \hat{\sigma}_z - \rho_{imp} \right) \, ,
\end{equation}
where $\tilde{\omega}_0 = \omega_0 + g \, n_0$ is the impurity energy splitting renormalized by the mean local density of the BH bath $n_0$. As anticipated before, the dephasing rate $\gamma{\left( t \right)}$ is completely determined by the time-dependent correlations of the bath operator coupled to the impurity -- local density fluctuations in the present case --
\begin{equation}\label{gamma}
\gamma{\left( t \right)} = \text{Re} \int_{0}^{t} d\tau \, \langle \hat{n}_{\mathbf{0}}{\left( \tau \right)} \, \hat{n}_{\mathbf{0}}{\left( 0 \right)} \rangle \, ,
\end{equation}
where we have defined $\langle \cdots \hspace{-0.08mm} \rangle = \textmd{Tr}{\left\{ \cdots \rho^0_{BH} \right\}}$.
We recall here that the derivation of \eqref{master_equation} does not require any assumption about the statistical properties of the environment, so that in principle the rate \eqref{gamma} can account also for weak-coupling effects of non-Gaussian correlations. Now, we highlight that the integrated rate 
\begin{equation}
\Gamma{\left( t \right)} = \int_{0}^{t} d\tau \, \gamma{\left( \tau \right)},
\label{Ggamma}
\end{equation}
is key to understanding the dephasing dynamics, as it establishes a direct connection between the decay rate $\gamma{\left( t \right)}$ and the physical consequences of its non-Markovian features.

In the framework of the open quantum system formalism, Breuer, Laine and Piilo (BLP) have proposed a rigorous definition for non-Markovianity of a generic quantum channel \cite{BLP}. Indeed, for the dephasing model studied in this work,
the BLP non-Markovianity measure depends directly on the decoherence function $\Gamma{\left( t \right)}$ via the so-called \textit{Loschmidt echo} \cite{hall_2014, unified_picture}
\begin{equation}
L{\left( t \right)} = \exp{\left[ -2 \left( g/\hbar \right)^2 \Gamma{\left( t \right)} \right]},
\label{echo}
\end{equation}
driving the off-diagonal evolution of the impurity state $\rho_{imp}{\left( t \right)}$~\footnote{See \ref{app_open_theory} for a detailed definition of the BLP non-Markovianity measure and its calculation in the pure dephasing model considered in this paper.}. In particular, the amount of non-Markovianity corresponds to the \textit{information back-flow} \cite{haikka2011, haikka_2012, haikka2013}
\begin{equation}\label{N}
\mathcal{N}_- = \sum_i \left[ \sqrt{L{\left( t_{i + 1} \right)}} - \sqrt{L{\left( t_i \right)}} \right] \, ,
\end{equation}
where the sum is taken over the set of time intervals $[t_i, t_{i + 1}]$ in which the echo increases, i.e. when $\gamma{\left( t \right)} < 0$.
During these intervals, some of the previously lost information regarding the state of the impurity is temporarily recovered. Conversely, the Markovian character of the dynamics $\mathcal{N}_+$ is quantified by summing $\sqrt{L{\left( t_{i + 1} \right)}} - \sqrt{L{\left( t_i \right)}}$ over the time intervals in which quantum information is lost. It is worth underlining that, for the special open quantum system that we consider here, all non-Markovianity measures agree in distinguishing Markovian from non-Markovian evolution \cite{rev_mod, apollaro}.

In the following sections, we will describe how a non-Markovian dephasing dynamics emerges due to strong correlations in the BH environment, focusing on the role of the universality classes of the MI-SF transition and on the importance of including non-Gaussian correlations beyond linear coupling between the bath excitations and the impurity (two-particle contributions). In this regard, we start our analysis by illustrating how the QGW approach provides semi-analytical expressions for the dephasing rate $\gamma{\left( t \right)}$ and the decoherence function $\Gamma{\left( t \right)}$, with a clear distinction between single-particle and non-Gaussian correlations.

\subsection{QGW expressions of \texorpdfstring{$\gamma{\left( t \right)}$}{g(t)} and \texorpdfstring{$\Gamma{\left( t \right)}$}{G(t)} and short-time behaviour of the Loschmidt echo \texorpdfstring{$L{\left( t \right)}$}{L(t)}}\label{app_explicit_gammas}
In this section we report for completeness the explicit expressions of the relevant quantities introduced above within the QGW formalism. Inserting the expression of the density operator \eqref{density_expansion} into the definition of the dephasing rate $\gamma{\left( t \right)}$, we can distinguish two contributions $\gamma{\left( t \right)}  = \gamma_1{\left( t \right)} + \gamma_2{\left( t \right)}$.
The first term is due to the linear-order part of the density operator $\eqref{BH_density_1}$,
\begin{equation}\label{gamma_1}
\gamma_1{\left( t \right)} = \text{Re} \int_{0}^{t} d\tau \left\langle \delta_1  \hat{n}_{\mathbf{0}}{\left( \tau \right)} \, \delta_1 \hat{n}_{\mathbf{0}}{\left( 0 \right)} \right\rangle = \frac{1}{V} \mathlarger\sum_{\alpha} \mathlarger\sum_{\mathbf{k}} N^2_{\alpha, \mathbf{k}} \frac{\sin{\left( \omega_{\alpha, \mathbf{k}} \, t \right)}}{\omega_{\alpha, \mathbf{k}}} \, ,
\end{equation}
while the second contribution is generated by the two-particle density operator \eqref{BH_density_2}, in particular
\begin{equation}\label{gamma_2}
\small
\gamma_2{\left( t \right)} = \text{Re} \int_{0}^{t} d\tau \left\langle \delta_2  \hat{n}_{\mathbf{0}}{\left( \tau \right)} \, \delta_2 \hat{n}_{\mathbf{0}}{\left( 0 \right)} \right\rangle = \frac{1}{V^2} \mathlarger\sum_{\alpha, \beta} \mathlarger\sum_{\mathbf{k}, \mathbf{p}} \left( W^2_{\alpha \mathbf{k}, \beta \mathbf{p}} + W_{\alpha \mathbf{k}, \beta \mathbf{p}} \, W_{\beta \mathbf{p}, \alpha \mathbf{k}} \right) \frac{\sin{\left[ \left( \omega_{\alpha, \mathbf{k}} + \omega_{\beta, \mathbf{p}} \right) \, t \right]}}{\omega_{\alpha, \mathbf{k}} + \omega_{\beta, \mathbf{p}}}
\end{equation}
at zero temperature. Analogously, the decoherence function is given by $\Gamma{\left( t \right)} = \Gamma_1{\left( t \right)} + \Gamma_2{\left( t \right)}$ with $\Gamma_i{\left( t \right)} = \int_{0}^{t} d\tau \, \gamma_i{\left( \tau \right)}$, $i=1, 2$.

The off-diagonal elements of the impurity density matrix will evolve according to the Loschmidt echo $L{\left( t \right)} = \exp{\left[ -2 \, \left( g/\hbar \right)^2 \, \Gamma{\left( t \right)} \right]} = \exp{\left[ -2 \left( g/\hbar \right)^2 \Gamma_1{\left( t \right)} \right]} \, \exp{\left[ -2 \left( g/\hbar \right)^2 \Gamma_2{\left( t \right)} \right]}$. From Eqs. \eqref{gamma_1}-\eqref{gamma_2} we see that the expected short-time Gaussian behaviour $\exp{\left[ -\lambda \left( g/\hbar \right)^2 t^2 \right]}$ \cite{Peres84} of the Loschmidt echo is recovered with
\begin{equation}\label{lambda}
\lambda = \frac{1}{V} \mathlarger\sum_{\alpha} \mathlarger\sum_{\mathbf{k}} N^2_{\alpha, \mathbf{k}} + \frac{1}{V^2} \mathlarger\sum_{\alpha, \beta} \mathlarger\sum_{\mathbf{k}, \mathbf{p}} \left( W^2_{\alpha \mathbf{k}, \beta \mathbf{p}} + W_{\alpha \mathbf{k}, \beta \mathbf{p}} \, W_{\beta \mathbf{p}, \alpha \mathbf{k}} \right).
\end{equation}
In the following we show how both $\lambda$ and the BLP non-Markovianity measure are not only extremely sensitive to the phase transition points, but behave differently depending on the universality class of the phase transition.

\begin{figure}[!pt]
    \centering
	\includegraphics[width=0.84\linewidth]{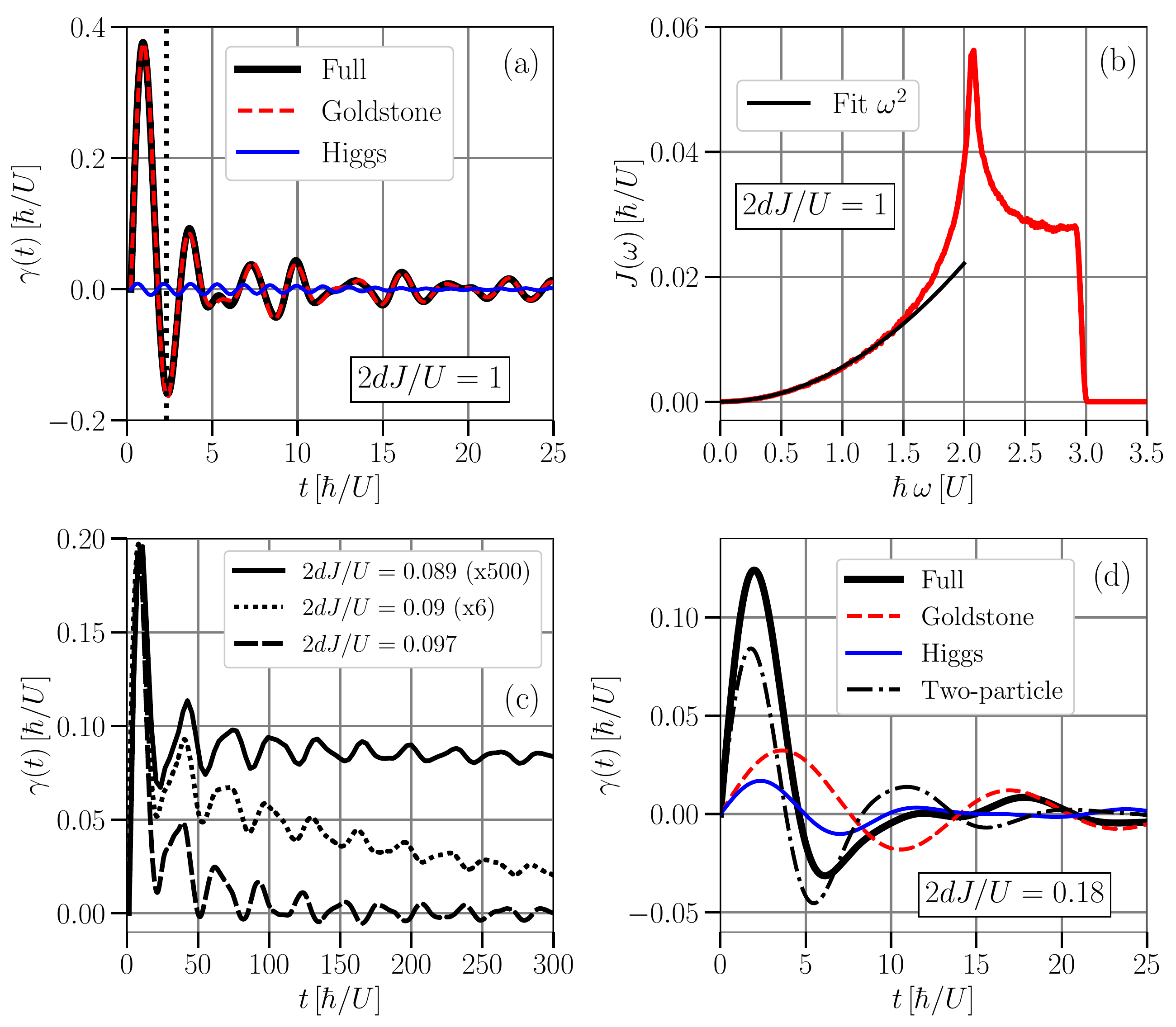}
    \caption{(a) Black solid line: dephasing rate $\gamma{\left( t \right)}$ at $2 \, d \, J/U = 1$ and $\mu/U = 0.8$ in the deep SF phase. Red dashed and blue solid lines: one-particle contributions of the Goldstone and Higgs modes respectively. The vertical black dashed line highlights the time scale $\tau_G$. (b) Red points: sampling of the spectral density $J{\left( \omega \right)}$ given by Eq.~\eqref{spectral} for $2 \, d \, J/U = 1$ and $\mu/U = 0.8$. Black solid line: $\omega^2$ fit of $J{\left( \omega \right)}$ at low $\omega$. (c) Change of $\gamma{\left( t \right)}$ while approaching the edge transition in the SF phase at $\left( 2 \, d \, J/U \right)^{edge}_c = 0.0\overline{8}$, with decreasing $2 \, d \, J/U$ from bottom to top. Magnification of $\gamma{\left( t \right)}$ at lower $2 \, d \, J/U$ is applied. (d) Black solid line: $\gamma{\left( t \right)}$ at $2 \, d \, J/U = 0.18$ and $\mu/U = \sqrt{2} - 1$, close to the tip critical point $\left( 2 \, d \, J/U \right)^{tip}_c \approx 0.172$ in the SF phase. Black dashed-dotted line: fraction of $\gamma{\left( t \right)}$ given by two-particle contributions involving the Goldstone and Higgs modes. The color code for the single-particle contributions is the same as in panel (a).}
    \label{SF}
\end{figure}

\section{Numerical results}\label{sec_results}

In the following we present the numerical results obtained by computing the dephasing rate functions \eqref{gamma_1}-\eqref{gamma_2} and the Loschmidt echo $L{\left( t \right)}$. All the calculations have been performed on a $400 \times 400$ square lattice, which well approximates the thermodynamic limit and is made possible by the low numerical complexity of the QGW approach. Moreover, we have imposed periodic boundary conditions so as to make the dephasing dynamics independent of the specific position of the impurity in the BH environment and avoid boundary effects. For brevity, hereafter we will refer to the CI transition as edge transition, while the O(2) critical point will be indicated as tip transition.

\subsection{Dephasing in the superfluid phase}\label{subsec_sf_dephasing}

We start our analysis about the dephasing dynamics starting from the weakly-interacting limit (deep SF phase) of the BH bath. In \autoref{SF}(a) we report the behaviour of the dephasing rate function $\gamma{\left( t \right)}$ [black solid line] for $2 \, d \, J/U = 1$. As expected, in this regime the contribution from the single-particle gapless Goldstone mode [red dashed line] saturates the time evolution of $\gamma(t)$. The dephasing rate exhibits broad oscillations around zero at short times, signalling the occurrence of non-Markovian effects, simply due to the finite bandwidth of the model. Very small amplitude oscillations persist at long times, leading to an essentially constant $\Gamma{\left( t \right)}$ and therefore only to a partial decoherence of the impurity density matrix.
For the sake of clarity, we argue a little bit on such result, which can be better understood by expressing the dephasing $\gamma{\left( t \right)} = \int_{0}^{\infty} d\omega \, J{\left( \omega \right)} \, \sin{\left( \omega \, t \right)}/\omega$ \cite{weiss2011} in terms of the single-particle spectral density 
\begin{equation}
J{\left( \omega \right)} = \sum_{\alpha, \mathbf{k}} N^2_{\alpha, \mathbf{k}} \, \delta{\left( \omega - \omega_{\alpha, \mathbf{k}} \right)} \, .
\label{spectral}
\end{equation}
This quantity for $2 \, d \, J/U = 1$ is shown in \autoref{SF}(b).
Being the Goldstone spectrum gapless and linear at small momenta, the spectral density scales as $J{\left( \omega \right)} \sim \omega^d$ at low frequencies~\footnote{We refer the reader to \ref{app_boson_baths} for an analytical derivation of the low-frequency scaling of $J{\left( \omega \right)}$ in the deep SF phase.}.
Nevertheless, in contrast with the non-Markovianity criterion generally adopted -- obtained in \cite{haikka2013} and fixing to $d > 2$ the necessary condition for memory effects in gapless baths --, we observe that $\gamma{\left( t \right)}$ has negative values in our $d = 2$ model. The reason is that usually an environment with infinite-bandwidth modes is considered in the literature \cite{haikka2013}, resulting in a smooth cutoff of the spectral density. The finite bandwidth of the BH model excitations implies a sharp frequency cutoff of $J{\left( \omega \right)}$ corresponding to the Goldstone mode energy at the edge of the Brillouin zone, $\omega_{G, \mathbf{\pi}}$. Correspondingly, we observe that the oscillations of $\gamma{\left( t \right)}$ occur on a time scale $\tau_G = 2 \, \pi/\omega_{G, \mathbf{\pi}}$ [vertical dotted line in \autoref{SF}(a)] set by the bandwidth of the Goldstone excitation~\footnote{See \ref{app_boson_baths} for an extensive discussion on the difference between lattice and continuous models at the level of the spectral density $J{\left( \omega \right)}$ and the dephasing function $\gamma{\left( t \right)}$.}.

\subsection{Dephasing dynamics at the MI-SF transition}\label{subsec_MI-SF}

Moving away from the deep SF phase and approaching the MI-SF critical region, the fate of the SF non-Markovian dynamics turns out to strongly depend on the type of crossed critical point.
In particular, crossing the edge transition [blue dashed line in \autoref{phase_diagram}] the amplitude of memory effects decreases with increasing interaction $U/J$ until the dynamics becomes purely Markovian on the Mott boundary. On the contrary, crossing the tip transition [red dashed line in \autoref{phase_diagram}(a)], the non-Markovianity is even more enhanced by quantum fluctuations with respect to the deep SF phase.

In panel (c) of \autoref{SF} we display the evolution of $\gamma{\left( t \right)}$ for different values of $2 \, d \, J/U$ upon approaching the edge transition. We observe that, close to the critical point $\left( 2 \, d \, J/U \right)^{edge}_c = 0.0\overline{8}$, $\gamma{\left( t \right)}$ becomes strictly positive and the dynamics slows down significantly, when compared with the evolution in the deep SF regime shown in panel (a). Therefore, at the edge critical point the dephasing rate reaches a constant value $\gamma{\left( t \right)} \sim \eta$ at asymptotically large times.
Hence, a transition from a non-Markovian to a Markovian regime occurs and, at the transition point, the Loschmidt echo acquires the typical exponential behaviour $L{\left( t \right)} \sim \exp{\left( -2 \, \eta \, g^2 \, t \right)}$ of a Lindbladian evolution. The origin of the Markovian behaviour is due to the peculiar spectral properties of the BH model on the edge of the Mott lobe. In particular, as illustrated in Subsection~\ref{subsec_qga}: (i) the Goldstone mode turns into an effective quasiparticle branch with quadratic energy dispersion; (ii) the Higgs mode keeps a finite energy gap.
It follows that the strongly-correlated superfluid sitting close to the edge critical point can be described as a dilute free-boson gas with an effective mass renormalized by the vicinity of the Mott phase \cite{sachdev, caleffi2019quantum}. Indeed, it is easy to check that for a free Bose gas -- with or without a lattice -- the Loschmidt echo decays always exponentially as $L{\left( t \right)} \sim e^{-\beta \, t}$ for $d = 2$~\footnote{We refer again the reader to \ref{app_boson_baths} for the explicit expressions of $\gamma{\left( t \right)}$ and $L{\left( t \right)}$ of a free boson gas on a lattice and on the continuum. See also \autoref{fig_SM_D} for exact numerical results on the behaviour of $\gamma{\left( t \right)}$ for lattice free bosons in one and two dimensions.}.
As in the deep SF case, the Goldstone single-particle contribution to $\gamma{\left( t \right)}$ is the dominant one, but, in this case, the two-particle contributions to $\gamma{\left( t \right)}$ are non-negligible in the edge critical region. However, we find that such a contribution integrates to zero identically in the time integral of the decoherence function $\Gamma{\left( t \right)} = \int_{0}^{t} d\tau \, \gamma{\left( \tau \right)}$. In this respect, the irrelevance of non-Gaussian bath correlations can be seen as a natural consequence of the effective single-particle description when crossing the CI critical region.

The result is very different when approaching the commensurate transition at the tip of the Mott lobe, as shown in panel (d). The dynamics appears to be always non-Markovian and the memory effects are amplified with respect to the deep SF regime. The dephasing rate $\gamma{\left( t \right)}$ gets a relevant contribution from the Higgs excitation and, most importantly, from the two-particle couplings [black dot-dashed line].
Specifically, the competition between the Goldstone and Higgs branches is evidently due to the closing of the Higgs gap at the tip critical point. For the same reason, one gets a sizable contribution to the dynamics from two-particle correlations due to the coupling between the Goldstone and Higgs modes encoded in the structure factors $W_{G, \mathbf{k}; H, \mathbf{p}}$ and $W_{H, \mathbf{p}; G, \mathbf{k}}$ in the two-mode part of the density operator \eqref{BH_density_2}.
Decreasing further $2 \, d \, J/U$ towards the critical point, non-Gaussian correlations eventually become the dominant contribution to $\gamma{\left( t \right)}$, since the order of magnitude of the single-particle weights $N_{\alpha, \mathbf{k}}$ is totally suppressed on the brink of the MI-SF transition \cite{caleffi2019quantum}.

In this respect, we want to stress that two-particle processes become the only non-vanishing contributions to density correlations when the BH environment enters the MI phase \cite{caleffi2019quantum}. Therefore, the dephasing dynamics undergoes a substantial change across the edge transition, where the single-particle picture is abruptly replaced by non-Gaussian correlations, while at the tip transition the single-to-two particle transfer of spectral weight appears to be a smoother crossover.

\subsection{Short-time dephasing process and non-Markovianity measure}\label{subsec_L}

\begin{figure}[!pt]
    \centering
	\includegraphics[width=0.9\linewidth]{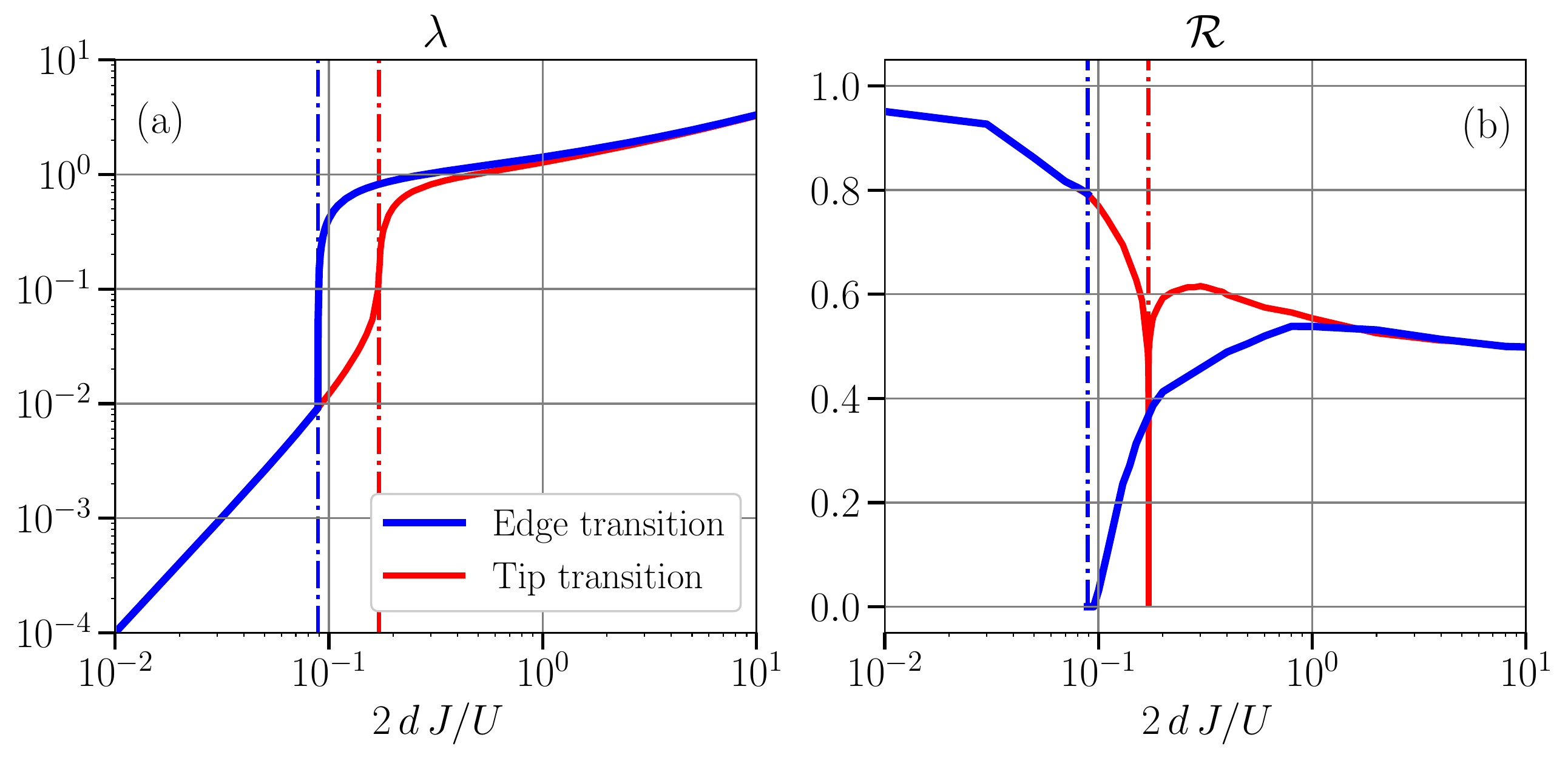}
    \caption{Panel (a): short-time decoherence rate $\lambda$ as a function of the rescaled hopping energy $2 \, d \, J/U$ across the edge (blue line) and tip (red line) transition points [see the phase diagram cuts in panel (a) of \autoref{phase_diagram}]. Panel (b): normalized information back-flow $\mathcal{R} = \mathcal{N}_-/\mathcal{N}_+$ for the same parameters. In both panels, the CI and O(2) critical points are indicated by blue and red dashed-dotted lines respectively.}
    \label{fig3}
\end{figure}

A concise way to visualize the previous results is provided by inspecting the dephasing dynamics from the point of view of the Loschmidt echo. Specifically, we focus our analysis on two complementary features of the decoherence process, namely (i) the short-time behaviour of the impurity decoherence $L{\left( t \to 0 \right)} = \exp{\left( -\lambda \, g^2 \, t^2 \right)}$ and (ii) the estimation of the information back-flow $\mathcal{N}_-$. 
More precisely, we renormalize the information back-flow by the overall coherence loss as $\mathcal{R} = \mathcal{N}_-/\mathcal{N}_+$, which provides a more effective measure of non-Markovianity while changing the bath parameters \cite{cosco_fermi}.

Our numerical results for the short-time decoherence rate $\lambda$, given by the expression \eqref{lambda}, are reported in panel (a) of \autoref{fig3}. Reaching the MI-SF critical region from the deep SF phase, the decoherence rate $\lambda$ decreases as a consequence of the stronger non-Markovianity driven by interactions in the BH bath. Reducing further the hopping energy, we observe that $\lambda$ presents different behaviours depending on the type of approached transition. At the CI critical points, the decoherence rate quickly drops to a small value (decreasing by almost two orders of magnitude) entering the MI phase, where we find that $\lambda \propto \left(J/U \right)^2$. The first derivative of $\lambda$ with respect to $J/U$ presents a discontinuity at the critical point.
Conversely, when crossing the transition at the lobe's tip, $\lambda$ is a smooth function of the hopping energy.
We notice that our latter result nicely resembles what has been found for the impurity decoherence process in a $d = 1$ interacting quantum spin bath \cite{fazio2006}, which has a critical point of the same $O(2)$ universality class. Therefore, as for the static properties \cite{caleffi2019quantum}, our method is able to capture the strong correlation also in this time dependent scenario, importantly beyond the one-dimensional case and without strong numerical requirements. 



The time-integrated dephasing dynamics, in the form of the non-Markovianity measure $\mathcal{R}$, is even more affected by the type of critical correlations than the short-time decoherence. Our numerical results for $\mathcal{R}$ across the edge and tip transitions are reported in panel (b) of \autoref{fig3} with the same color code of panel (a). In particular, for the calculation of $\mathcal{R}$ we have fixed $g/U = 0.001 \ll 2 \, d \, J/U, \mu/U$ coherently with the weak-coupling condition.

In the deep SF limit $J/U \gg 1$, we find that both the information flows $\mathcal{N}_{\pm}$ tend to zero scaling as $\left( J/U \right)^{-1}$, such that their ratio $\mathcal{R}$ is a constant. This indicates that, when embedded in a weakly-interacting gas, the impurity dephases according to a fixed fraction of information loss.
When approaching the strongly-interacting regime, the renormalized back-flow $\mathcal{R}$ reaches a maximal value well before the MI-SF transition. This suggests that, away from critical region, the primary effect of stronger interactions is to increase the amount of information recovered by the impurity during the dynamics. When approaching the critical point the non-Markovianity measure $\mathcal{R}$ starts decreasing and its behaviour depends on how the MI-SF is crossed.

Crossing the CI transition $\mathcal{R}$ rapidly vanishes, being zero within a small window in the SF region.
This result perfectly mirrors the non-Markovian to Markovian transition displayed in \autoref{SF}(c) and the effective free-particle description of the SF at the CI critical points.
The quantity $\mathcal{R}$ show a discontinuous behaviour, when entering the insulating phase. This result finds a straightforward interpretation in terms of the particle-hole excitations of the Mott phase \cite{cosco2018}.
Due to their incoherent character, these modes excite doublon-holon pairs with a finite correlation length, so that density fluctuations are localized in real space. Therefore, when particle-hole excitations couple to the impurity, the information flowing to the BH environment remains localized in a small neighbourhood of the impurity and is likely to be restored after a short time due to another particle-hole excitation. As the amplitude of density fluctuations in the Mott phase increase with $2 \, d \, J/U$, the absolute value of both the information flows $\mathcal{N}_{\pm}$ increases accordingly; on the other hand, the renormalized back-flow $\mathcal{R}$ decrease as a consequence of the increasing BH correlation length, which prevent part of the lost information from flowing back to the impurity. However, since at the edge transition either the particle or the hole branch remains gapped, a finite correlation length still controls the dynamics exactly at the critical point \cite{sachdev, caleffi2019quantum}, before diverging in the SF phase. This discontinuous behaviour of the correlation length is at the roots the finite jump in $\mathcal{R}$ across the non-Markovian to Markovian transition.

The behaviour is different at the tip transition. As shown before, in this regime critical fluctuations are mainly due to non-Gaussian correlations, whose main effect is to amplify the oscillation amplitude of the dephasing rate $\gamma{\left( t \right)}$. Therefore, the amount of total information flowing both from and to the impurity grows accordingly. Nevertheless, the renormalized backflow $\mathcal{R}$ still converges to zero at the critical point, meaning that eventually the BH environment becomes effectively Markovian at the critical point. It follows that, in contrast with the edge case, $\mathcal{R}$ is found to be a continuous function of the hopping $2 \, d \, J/U$ across the tip transition, but with a very sharp non-monotonic profile
[red line in \autoref{fig3}(b)]. 



\section{Summary and outlook}\label{sec_summary}

In this paper, we present an exhaustive account of the non-Markovian effects characterizing the dephasing dynamics of an impurity embedded in a Bose-Hubbard environment undergoing the superfluid-Mott transition. 

Our analysis addresses the impurity problem beyond the standard formalism of open quantum systems. The two main new features are the inclusion of the effects of the strong correlations and phase transitions in the environment and the extension beyond the one-dimensional case in a flexible and numerically cheap way. Thereby, our method is, to the best of our knowledge, the first one that allows an efficient description of an open system that is coupled to an environment undergoing a critical transition.


Strong signatures of deviation from a Markovian behavior due to the spatial discreteness of the lattice setup, not explicitly discussed in previous works, have also been highlighted in the interacting superfluid phase and related to key features of the spectral density $J{\left( \omega \right)}$. This suggests the idea that the very same phenomenon could take place in different lattice models whose dynamics is governed by common spectral properties. Furthermore, we observed that the amount of non-Markovianity of the dephasing process is particularly large when approaching the O(2) critical region, where two-particle effects become more relevant in the physical picture and thus the environment differs more significantly from the standard spin-boson description. This opens the path for further investigations into the role of strong non-Gaussian, i.e. two-particle, correlations in the presence of strong memory effects.

More importantly, we have found that, when the BH environment approaches the superfluid-Mott criticality, the dephasing dynamics is extremely sensitive to the universality class of the superfluid-Mott transition. In this regard, we have shown that not only the deviation from Markovianity, but also the short-time behaviour of the dephasing dynamics carries strong signatures of the type of criticality approached by the environment. This remarkable result
agrees with similar findings for interacting quantum spin baths \cite{fazio2006} with a complementary approach, suggesting a generality which goes beyond the precise nature and the dimensionality of the bath.




Finally, from an experimental perspective, the sharp difference between the dephasing processes at the different superfluid-Mott transitions discussed in this work identifies the study of the decoherence dynamics and, in particular, non-Markovianity measures of impurity dephasing as an unambiguous probe of the type of critical behaviour experienced by the environment.

\vspace{1cm}

\section*{Acknowledgements}
Financial support from the Italian MIUR under the PRIN2017 project CEnTraL (Protocol Number 20172H2SC4) and from the Provincia Autonoma di Trento is acknowledged. I.d.V. was  financially supported by DFG-Grant GZ: VE 993/1-1.

\pagebreak

\renewcommand\appendixname{APPENDIX}
\appendix
\renewcommand{\theequation}{\thesection.\arabic{equation}}

\section{Quantum Gutzwiller approach in a nutshell}\label{app_qga}
The QGW approach combines the successful features of the Gutzwiller approximation \cite{gutzwiller_1991} and the Bogoliubov theory of weakly-interacting gases \cite{rey} in order to develop a robust quantum many-body theory of a generic interacting lattice model. Building on the solution of the time-dependent Gutzwiller approximation \cite{krutitsky}, fluctuations on top of the mean-field ground state are quantized in terms of the elementary many-body excitations of the system and systematically included in the calculation of ground state expectation values. In spite of the local nature of the underlying Gutzwiller ansatz -- see Eq.~\eqref{Gutz_ansatz} below --, the QGW approach accurately reproduce both local and non-local correlations across the different phases of the BH model with minimal numerical effort, showing a remarkable agreement with quantum Monte Carlo predictions concerning density correlations.
Let us also mention that the QGW, when only quadratic fluctuations are considered, coincides essentially with including quantum fluctuations by slave boson approaches (see in particular \cite{roscilde}, where the slave boson approach has been applied to the BH Hamiltonian to determine its entanglement entropy along its phase diagram).

Following the main derivation steps of \cite{caleffi2019quantum}, in this Appendix we briefly review the essential features of QGW technique, that we employ for a systematic evaluation of quantum correlations in the BH environment.

Our starting point is the Gutzwiller ansatz
\begin{SEquation}\label{Gutz_ansatz}
| \Psi_G \rangle = \bigotimes_{\mathbf{r}} \sum_n c_n{\left( \mathbf{r} \right)} \, | n, \mathbf{r} \rangle \, ,
\end{SEquation}
where the wave function is site-factorized
and the complex amplitudes $c_n{\left( \mathbf{r} \right)}$ of each local Fock state $| n, \mathbf{r} \rangle$  are variational parameters with normalization condition $\sum_n \left| c_n{\left( \mathbf{r} \right)} \right|^2 = 1$. In our specific case, we draw on the simple form of \eqref{Gutz_ansatz} to reformulate the BH model in terms of the following Lagrangian functional
\begin{SEquation}\label{lagrangian}
\begin{aligned}
\mathfrak{L}{\left[ c, c^* \right]} &= \big\langle \Psi_G \big| \, i \, \hbar \, \partial_t - \hat{H}_{BH} \, \big| \Psi_G \big\rangle \\
&= \frac{i \, \hbar}{2} \sum_{\mathbf{r},n}  [c^{*}_n(\mathbf{r}) \dot{c}_n(\mathbf{r}) - \textrm{c.c.}] + J \sum_{\langle \mathbf{r}, \mathbf{s} \rangle} \left[ \psi^{*}{\left( \mathbf{r} \right)} \, \psi{\left( \mathbf{s} \right)} + \text{c.c.} \right] - \sum_{\mathbf{r},n} H_n \left| c_n{\left( \mathbf{r} \right)} \right|^2 \, .
\end{aligned}
\end{SEquation}
In the previous equation, the dot indicates the temporal derivative,
\begin{SEquation}
H_n = \frac{U}{2}  n \left( n - 1 \right) - \mu \, n
\end{SEquation}
are the matrix elements of the on-site terms of the BH Hamiltonian $\hat{H}_{BH}$ in Fock space and
\begin{SEquation}
\psi{\left( \mathbf{r} \right)}=
\big\langle \hat{a}_{\mathbf{r}} \big\rangle = \sum_n \sqrt{n} \, c^*_{n - 1}{\left( \mathbf{r} \right)} \, c_n{\left( \mathbf{r} \right)}
\end{SEquation}
is the mean-field order parameter. In this formulation, the conjugate momenta of the parameters $c_n{\left( \mathbf{r} \right)}$ are $c^*_n{\left( \mathbf{r} \right)} = \partial \mathfrak{L}/\partial \dot{c}_n{\left( \mathbf{r} \right)}$. The classical Euler-Lagrange equations associated to Lagrangian \eqref{lagrangian} are the so-called \textit{time-dependent Gutzwiller equations} as derived, e.g., in \cite{sheshandri_krishnamurthy_pandit_ramakrishnan, krutitsky_navez}. In a uniform system, the stationary solutions are homogeneous: in particular, the system is in a Mott Insulator (MI) state if $\psi{\left( \mathbf{r} \right)} = 0$ and in a superfluid (SF) state otherwise.

In order to go beyond the Gutzwiller approximation introduced above, it is natural to consider how quantum effects populate the excitation modes of the system and to investigate how they affect the observable quantities. We include quantum fluctuations by building a theory of the excitations starting from Lagrangian \eqref{lagrangian} via canonical quantization \cite{cohen, blaizot_ripka}, namely promoting the coordinates of the theory and their conjugate momenta to operators and imposing equal-time canonical commutation relations
\begin{SEquation}
\left[ \hat{c}_n{\left( \mathbf{r} \right)}, \hat{c}^{\dagger}_m{\left( \mathbf{s} \right)} \right] = \delta_{\mathbf{r},\mathbf{s}} \, \delta_{n,m} \, .
\end{SEquation}
In analogy with the Bogoliubov approximation for dilute Bose-Einstein condensates \cite{pitaevskii_stringari, castin}, we expand the operators $\hat{c}_n$ around their ground state values $c^0_n$, obtained by minimizing the energy $\big\langle \Psi_G \big| \hat{H}_{BH} \big| \Psi_G \big\rangle$, as 
\begin{SEquation}
\hat{c}_n{\left( \mathbf{r} \right)} = \hat{A}{\left( \mathbf{r} \right)} \, c^0_n + \delta \hat{c}_n{\left( \mathbf{r} \right)} \, .
\end{SEquation}
The \textit{normalization operator} $\hat{A}{\left( \mathbf{r} \right)}$ is a function of $\delta \hat{c}_n\left( \mathbf{r} \right)$ and $\delta \hat{c}^{\dagger}_n\left( \mathbf{r} \right)$ and ensures the proper normalization $\sum_n \hat{c}^{\dagger}_n{\left( \mathbf{r} \right)} \, \hat{c}_n{\left( \mathbf{r}\right)} = \hat{\mathds{1}}$. By restricting to local fluctuations orthogonal to the ground state $\sum_n \delta \hat{c}^\dagger_n{\left( \mathbf{r} \right)} \, c^{0}_n  = 0$ one has
\begin{SEquation}
\hat{A}{\left( \mathbf{r} \right)} = \left[ 1 - \sum_n \delta \hat{c}^{\dagger}_n{\left( \mathbf{r} \right)} \, \delta \hat{c}_n{\left( \mathbf{r}\right)} \right]^{1/2} \, .
\end{SEquation}
In a homogeneous system, it is convenient to work in momentum space by writing
\begin{SEquation}\label{c_FT}
\delta \hat{c}_n{\left( \mathbf{r} \right)} \equiv V^{-1/2} \sum_{\mathbf{k} \in \text{BZ}} e^{i \mathbf{k} \cdot \mathbf{r}} \, \delta \hat{C}_n{\left( \mathbf{k} \right)} \, .
\end{SEquation}
where $V$ is the lattice volume. Inserting Eq.~\eqref{c_FT} in $\langle \Psi_G | \hat{H}_{BH} \big| \Psi_G \big\rangle$ and keeping only terms up to the quadratic order in the fluctuations, we obtain
\begin{SEquation}\label{H_QGA_unrotated}
\hat{H}^{\left( 2 \right)}_{QGW} = E_0 + \frac{1}{2} \sum_{\mathbf{k}}
[\delta \underline{\hat{C}}^{\dagger}{\left( \mathbf{k} \right)} , -\delta \underline{\hat{C}}{\left( -\mathbf{k} \right)}]
\,\hat{\mathcal{L}}_{\mathbf{k}}
\begin{bmatrix}
\delta \underline{\hat{C}}{\left( \mathbf{k} \right)} \\
\delta \underline{\hat{C}}^{\dagger}{\left( -\mathbf{k} \right)}
\end{bmatrix} \, ,
\end{SEquation}
where $E_0$ is the mean-field ground state energy, the vector $\delta \underline{\hat{C}}(\mathbf{k})$ contains the components $\delta{\hat{C}}_n(\mathbf{k})$, and $\hat{\mathcal{L}}_{\mathbf{k}}$ is a pseudo-Hermitian matrix, for the explicit expression of which we refer the interested reader to \cite{caleffi2019quantum}. A suitable Bogoliubov rotation of the Gutzwiller operators in terms of the fundamental excitation modes of the system
\begin{SEquation}\label{C_many_body}
\delta \hat{C}_n{\left( \mathbf{k} \right)} = \sum_{\alpha} u_{\alpha, \mathbf{k}, n} \, \hat{b}_{\alpha, \mathbf{k}} + \sum_{\alpha} v^{*}_{\alpha, -\mathbf{k}, n} \, \hat{b}^{\dagger}_{\alpha, -\mathbf{k}} \, ,
\end{SEquation}
recasts the quadratic form \eqref{H_QGA_unrotated} into the desired diagonal form
\begin{SEquation}\label{H_QGA_app}
\hat{H}^{\left( 2 \right)}_{QGW} = \sum_{\alpha} \sum_{\mathbf{k}} \omega_{\alpha, \mathbf{k}} \, \hat{b}^{\dagger}_{\alpha, \mathbf{k}} \hat{b}_{\alpha, \mathbf{k}} \, ,
\end{SEquation}
where each $\hat{b}_{\alpha, \mathbf{k}}$ corresponds to a different many-body excitation mode with frequency $\omega_{\alpha,\mathbf{k}}$, labeled by its momentum $\mathbf{k}$ and branch index $\alpha$. Bosonic commutation relations between the annihilation and creation operators $\hat{b}_{\alpha, \mathbf{k}}$ and  $\hat{b}^\dagger_{\alpha, \mathbf{k}}$,
\begin{SEquation}
\left[ \hat{b}_{\alpha, \mathbf{k}}, \hat{b}^{\dagger}_{\alpha', \mathbf{k'}} \right] = \delta_{\mathbf{k}, \mathbf{k}'} \, \delta_{\alpha, \alpha'} \, ,
\end{SEquation}
are enforced by choosing the usual Bogoliubov normalization condition 
\begin{SEquation}
\underline{u}^{*}_{\alpha, \mathbf{k}} \cdot \underline{u}_{\beta, \mathbf{k}} - \underline{v}^{*}_{\alpha, -{\mathbf{k}}} \cdot \underline{v}_{\beta, -\mathbf{k}} = \delta_{\alpha \beta} \, ,
\end{SEquation}
where the vectors $\underline{u}_{\alpha, \mathbf{k}}$ ($\underline{v}_{\alpha, \mathbf{k}}$) contain the components $u_{\alpha, \mathbf{k}, n}$ ($v_{\alpha, \mathbf{k}, n}$).

The effective, quadratic description of the BH environment in terms of its collective modes \eqref{H_QGA_app} provided the QGW not only allows for a direct reinterpretation of the pure dephasing model \eqref{model}, but also opens a simple route to the calculation of any expectation value of the bath operators. Based on the quantization procedure outlined before, the evaluation of average value of any observable $\big\langle \hat{O}{\left( \hat{a}^{\dagger}_{\mathbf{r}}, \hat{a}_{\mathbf{r}} \right)} \big\rangle$ consists in applying a four-step procedure that we summarize as follows:
\begin{enumerate}
    \item Determine the expression $\mathcal{O}{\left[ c, c^* \right]} = \big\langle \Psi_G \big| \hat{O} \big| \Psi_G \big\rangle$ in terms of the Gutzwiller parameters $c_n$ and $c^*_n$;
    \item Create the operator $\hat{\mathcal{O}}{\left[ \hat{c}, \hat{c}^{\dagger} \right]}$ by replacing the Gutzwiller parameters in $\mathcal{O}{\left[ c, c^* \right]}$ by the corresponding operators $\hat{c}_n{\left( \mathbf{r} \right)}$ and $\hat{c}^{\dagger}_n{\left( \mathbf{r} \right)}$ without modifying their ordering;
    \item  Expand the operator $\hat{\mathcal{O}}$ order by order in the fluctuations $\delta \hat{c}_n$ and $\delta \hat{c}^{\dagger}_n$, taking into account the dependence of the operator $\hat{A}$ on the fluctuation operators. The contribution of $\hat{A}$ may be of fundamental importance when higher orders in the fluctuations become relevant;
    \item Taking advantage of the quadratic character of the QGW Hamiltonian, invoke Wick theorem to compute the expectation value of products of operators on Gaussian states -- such as ground or thermal states obtained from $H^{\left( 2 \right)}_{QGW}$.
\end{enumerate}
The very same protocol determines the expansion of the BH local density operator \eqref{density_expansion} in terms of single \eqref{BH_density_1} and two-particle \eqref{BH_density_2} operator-valued expressions of the collective modes $\hat{b}_{\alpha, \mathbf{k}}$, from which the bath correlation functions are systematically extracted. For the sake of completeness, we report the exact expressions of the one and two-particle structure factors of the density channel,
\begin{SEquation}\label{structure_factors}
\begin{gathered}
N_{\alpha, \mathbf{k}} = \sum_n c^0_n \left( u_{\alpha, \mathbf{k}, n} + v_{\alpha, \mathbf{k}, n} \right) \\
W_{\alpha \mathbf{k}, \beta \mathbf{p}} = \sum_n \left( n - n_0 \right) u_{\alpha, \mathbf{k}, n} \, v_{\beta, \mathbf{p}, n} \\
U_{\alpha \mathbf{k}, \beta \mathbf{p}} = \sum_n \left( n - n_0 \right) u_{\alpha, \mathbf{k}, n} \, u_{\beta, \mathbf{p}, n} \\
V_{\alpha \mathbf{k}, \beta \mathbf{p}} = \sum_n \left( n - n_0 \right) v_{\alpha, \mathbf{k}, n} \, v_{\beta, \mathbf{p}, n}
\end{gathered}
\end{SEquation}
whose derivation is extensively discussed in \cite{caleffi2019quantum}.

\section{Pure dephasing and BLP non-Markovianity measure}\label{app_open_theory}
The definition of the Breuer-Laine-Piilo (BLP) measure \cite{BLP} derives from considering non-Markovian those systems in which a back-flow of information from the environment to the open system occurs during the dynamics. This information recovery is formally identified by an increase in the distinguishability of pairs of evolving quantum states of the system.

In detail, a system is non-Markovian if there is a pair of system initial states $\rho^{\left( 1 \right)}_S{\left( 0 \right)}$ and $\rho^{\left( 2 \right)}_S{\left( 0 \right)}$, such that for certain times $t > 0$ their distinguishability grows, namely
\begin{SEquation}\label{dist_sigma}
\sigma{\left[ \rho^{\left( 1 \right)}_S{\left( 0 \right)}, \rho^{\left( 2 \right)}_S{\left( 0 \right)}; t \right]} = \frac{d}{dt} \mathcal{D}{\left[ \rho^{\left( 1 \right)}_S{\left( t \right)}, \rho^{\left( 2 \right)}_S{\left( t \right)} \right]} > 0 \, ,
\end{SEquation}
where $\sigma{\left[ \rho^{\left( 1 \right)}_S, \rho^{\left( 2 \right)}_S; t \right]}$ is called the \textit{information flux} at time $t$ and
\begin{SEquation}\label{dist}
\mathcal{D}{\left[ \rho^{\left( 1 \right)}_S{\left( t \right)}, \rho^{\left( 2 \right)}_S{\left( t \right)} \right]} \doteq \frac{1}{2} \left|\left| \rho^{\left( 1 \right)}_S{\left( t \right)} - \rho^{\left( 2 \right)}_S{\left( t \right)} \right|\right|_1 = \frac{1}{2} \text{Tr}{\left\{ \sqrt{\left[  \rho^{\left( 1 \right)}_S{\left( t \right)} - \rho^{\left( 2 \right)}_S{\left( t \right)} \right]^{\dagger} \left[  \rho^{\left( 1 \right)}_S{\left( t \right)} - \rho^{\left( 2 \right)}_S{\left( t \right)} \right]} \right\}}
\end{SEquation}
is defined to be the distinguishability of $\rho^{\left( 1 \right)}_S$ and $\rho^{\left( 2 \right)}_S$. Since density matrices are Hermitian, we have that
\begin{SEquation}\label{dist_eigen}
\mathcal{D}{\left[ \rho^{\left( 1 \right)}_S{\left( t \right)}, \rho^{\left( 2 \right)}_S{\left( t \right)} \right]} = \frac{1}{2} \text{Tr}{\left\{ \sqrt{\left[  \rho^{\left( 1 \right)}_S{\left( t \right)} - \rho^{\left( 2 \right)}_S{\left( t \right)} \right]^2} \right\}} = \frac{1}{2} \sum_i \left| \lambda_i \right| \, ,
\end{SEquation}
where $\lambda_i$ are the eigenvalues of the matrix $\rho^{\left( 1 \right)}_S - \rho^{\left( 2 \right)}_S$. The physical interpretation of the trace distance \eqref{dist} is that it is related to the maximum probability of distinguishing between two quantum states. In an open quantum system, this probability in general tends to decrease in time, as the system information is lost to the environment, except when the dynamics is non-Markovian. In this case, the system regains part of the previously lost information. According to the BLP criterion, the amount of non-Markovianity of a quantum process $\Lambda$ can be quantified through the measure
\begin{SEquation}\label{BLP_measure}
\mathcal{N}_-{\left( \Lambda \right)} = \text{max}_{\rho_{1, 2}{\left( 0 \right)}} \mathlarger\int_{\sigma > 0} dt \, \sigma{\left[ \rho^{\left( 1 \right)}_S{\left( 0 \right)}, \rho^{\left( 2 \right)}_S{\left( 0 \right)}; t \right]} \, ,
\end{SEquation}
which reflects the maximum amount of information that can flow back to the system for a given process. As proven in \cite{wissmann}, for all finite-dimensional quantum systems the evaluation of \eqref{BLP_measure} can be optimized by considering initial states $\rho^{\left( 1 \right)}_S{\left( 0 \right)}$ and $\rho^{\left( 2 \right)}_S{\left( 0 \right)}$ that are orthogonal and lie on the boundary of the subset of physical states. \\
In the case of the two-level impurity undergoing pure dephasing studied in this paper, the open system dynamics is driven by the master equation \eqref{master_equation}, which allows for a simple rewriting in the vector representation of the density matrix,
\begin{SEquation}\label{vector_master_equation}
\frac{d}{dt}
\begin{pmatrix}
\rho_{11} \\
\rho_{12} \\
\rho_{21} \\
\rho_{22}
\end{pmatrix}
=
\begin{pmatrix}
0 & 0 & 0 & 0 \\
0 & -\left( g/\hbar \right)^2 \gamma{\left( t \right)} & 0 & 0 \\
0 & 0 & -\left( g/\hbar \right)^2 \gamma{\left( t \right)} & 0 \\
0 & 0 & 0 & 0
\end{pmatrix}
\begin{pmatrix}
\rho_{11} \\
\rho_{12} \\
\rho_{21} \\
\rho_{22}
\end{pmatrix}
\, ,
\end{SEquation}
where we have defined $\rho_{ij} = \text{Tr}_S{\left\{ \rho_S{\left( t \right)} \left| i \right\rangle \left\langle j \right| \right\}}$, with $\left| i \right\rangle = \left| 1 \right\rangle, \left| 2 \right\rangle$ standing for the two possible states of the impurity, and neglected the unitary evolution terms set by the renormalized transition frequency $\tilde{\omega}_0$. The analytical integration of \eqref{vector_master_equation} yields
\begin{SEquation}\label{vector_master_equation_solution}
\rho_S{\left( t \right)} = \phi_t{\left[ \rho_S{\left( 0 \right)} \right]} =
\begin{pmatrix}
\rho_{11}{\left( 0 \right)} & \rho_{12}{\left( 0 \right)} \, \sqrt{L{\left( t \right)}} \\
\rho_{21}{\left( 0 \right)} \, \sqrt{L{\left( t \right)}} & \rho_{22}{\left( 0 \right)}
\end{pmatrix}
\, ,
\end{SEquation}
where $\phi_t$ is the dynamical map of the system density matrix associated to the pure dephasing dynamics. The function
\begin{SEquation}\label{loschmidt_echo}
L{\left( t \right)} = \exp{\left[ -2 \left( g/\hbar \right)^2 \mathlarger\int_{0}^{t} d\tau \, \gamma{\left( \tau \right)} \right]}
\end{SEquation}
coincides with the so-called \textit{Loschmidt echo} \cite{unified_picture}, defined as $L{\left( t \right)} = \left| \langle \psi{\left( t \right)} | \psi_0{\left( t \right)} \rangle \right|^2$, where $| \psi_0{\left( t \right)} \rangle$ is the bath ground state evolved according to its own Hamiltonian, while $| \psi{\left( t \right)} \rangle$ is the time-evolved bath state in presence of the open system. Indeed, the off-diagonal matrix elements of the system density matrix $\rho_S$ are given by $\sqrt{L{\left( t \right)}}$ exactly. \\
Choosing two initial states that are orthogonal and lie on the Bloch sphere of the two-level system
\begin{SEquation}\label{initial_states}
\rho^{\left( 1 \right)}_S{\left( 0 \right)} = \frac{1}{2}
\begin{pmatrix}
1 & 1 \\
1 & 1
\end{pmatrix}
\qquad \rho^{\left( 2 \right)}_S{\left( 0 \right)} = \frac{1}{2}
\begin{pmatrix}
1 & -1 \\
-1 & 1
\end{pmatrix}
\, ,
\end{SEquation}
we find that the the trace distance \eqref{dist} reads
\begin{SEquation}\label{dist_dephasing}
\mathcal{D}{\left[ \rho^{\left( 1 \right)}_S, \rho^{\left( 2 \right)}_S \right]} = \frac{1}{2} \left|\left| \rho^{\left( 1 \right)}_S{\left( t \right)} - \rho^{\left( 2 \right)}_S{\left( t \right)} \right|\right|_1 = \left|\left|
\begin{pmatrix}
0 & \sqrt{L{\left( t \right)}} \\
\sqrt{L{\left( t \right)}} & 0
\end{pmatrix}
\right|\right|_1 = \sqrt{L{\left( t \right)}} \, .
\end{SEquation}
Therefore, we obtain that the distinguishability rate is given by
\begin{SEquation}\label{dist_sigma_dephasing}
\sigma{\left[ \rho^{\left( 1 \right)}_S, \rho^{\left( 2 \right)}_S; t \right]} = \frac{d\mathcal{D}{\left[ \rho^{\left( 1 \right)}_S, \rho^{\left( 2 \right)}_S \right]}}{dt} = -\left( g/\hbar \right)^2 \gamma{\left( t \right)} \, \sqrt{L{\left( t \right)}}
\end{SEquation}
and $\sigma{\left[ \rho^{\left( 1 \right)}_S, \rho^{\left( 2 \right)}_S; t \right]} > 0$ for some $t$ when the dephasing rate $\gamma{\left( t \right)}$ is negative, leading to non-Markovian dynamics. Finally, it is straightforward to deduce that the non-Markovianity measure \eqref{BLP_measure} is provided by the values of the Loschmidt echo $L{\left( t \right)}$ at the boundaries of those time intervals $\left[ t_i, t_{i + 1} \right]$ over which $\gamma{\left( t \right)} < 0$, namely
\begin{SEquation}\label{BLP_measure_dephasing}
\mathcal{N}_- = \mathlarger\int_{\sigma > 0} dt \, \sigma{\left[ \rho^{\left( 1 \right)}_S{\left( 0 \right)}, \rho^{\left( 2 \right)}_S{\left( 0 \right)}; t \right]} = -\mathlarger\int_{\gamma < 0} dt \, \left( g/\hbar \right)^2 \gamma{\left( t \right)} \, \sqrt{L{\left( t \right)}} = \sum_i \left[ \sqrt{L{\left( t_{i + 1} \right)}} - \sqrt{L{\left( t_i \right)}} \right]
\end{SEquation}
by the definition of $L{\left( t \right)}$. On an equal footing, we can also quantify the amount of information that flows from the open system to the environment by defining a Markovianity measure
\begin{SEquation}\label{anti_BLP_measure_dephasing}
\mathcal{N}_+ = \mathlarger\int_{\sigma < 0} dt \, \sigma{\left[ \rho^{\left( 1 \right)}_S{\left( 0 \right)}, \rho^{\left( 2 \right)}_S{\left( 0 \right)}; t \right]} = -\mathlarger\int_{\gamma > 0} dt \, \left( g/\hbar \right)^2 \gamma{\left( t \right)} \, \sqrt{L{\left( t \right)}} \, ,
\end{SEquation}
which takes into account time periods for which $\gamma{\left( t \right)} > 0$.

\section{Dephasing dynamics at incommensurate filling}\label{app_incommensurate}
\begin{figure}[!htp]
    \centering
	\includegraphics[width=0.45\linewidth]{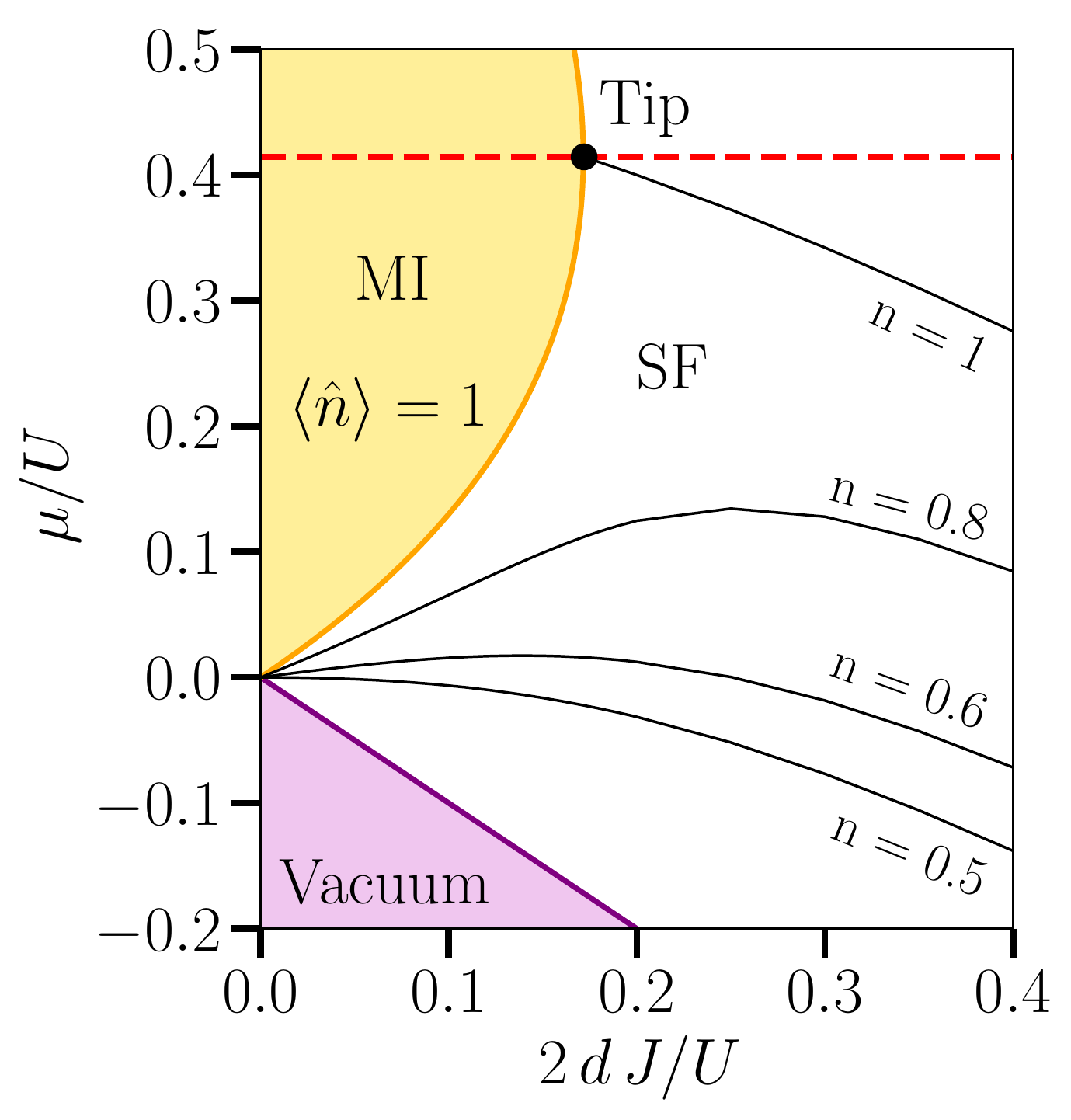}
    \caption{Detail of the mean-field phase diagram of the BH model [see panel (a) of \autoref{phase_diagram}] showing typical constant-density lines (black solid lines) in the SF phase. Non-integer filling lines connect the hard-core regime $\left( 2 \, d \, J/U \ll 1 \right)$ to the deep SF phase at $2 \, d \, J/U \gtrsim 1$.}
    \label{incommensurate_lines}
\end{figure}

In this Appendix, we report and discuss the quantitative evolution of the dephasing rate $\gamma{\left( t \right)}$ and of the Loschmidt echo $L{\left( t \right)}$ as the BH bath becomes strongly-interacting without entering the Mott phase and, on the contrary, retaining a superfluid character. Specifically, this corresponds to reach the hard-core boson limit of the BH model by increasing the boson interaction $U$ at fixed non-commensurate density. Typical constant-density contours in the strongly-interacting SF phase are shown in \autoref{incommensurate_lines}.

\begin{figure}[!htp]
    \centering
	\includegraphics[width=0.95\linewidth]{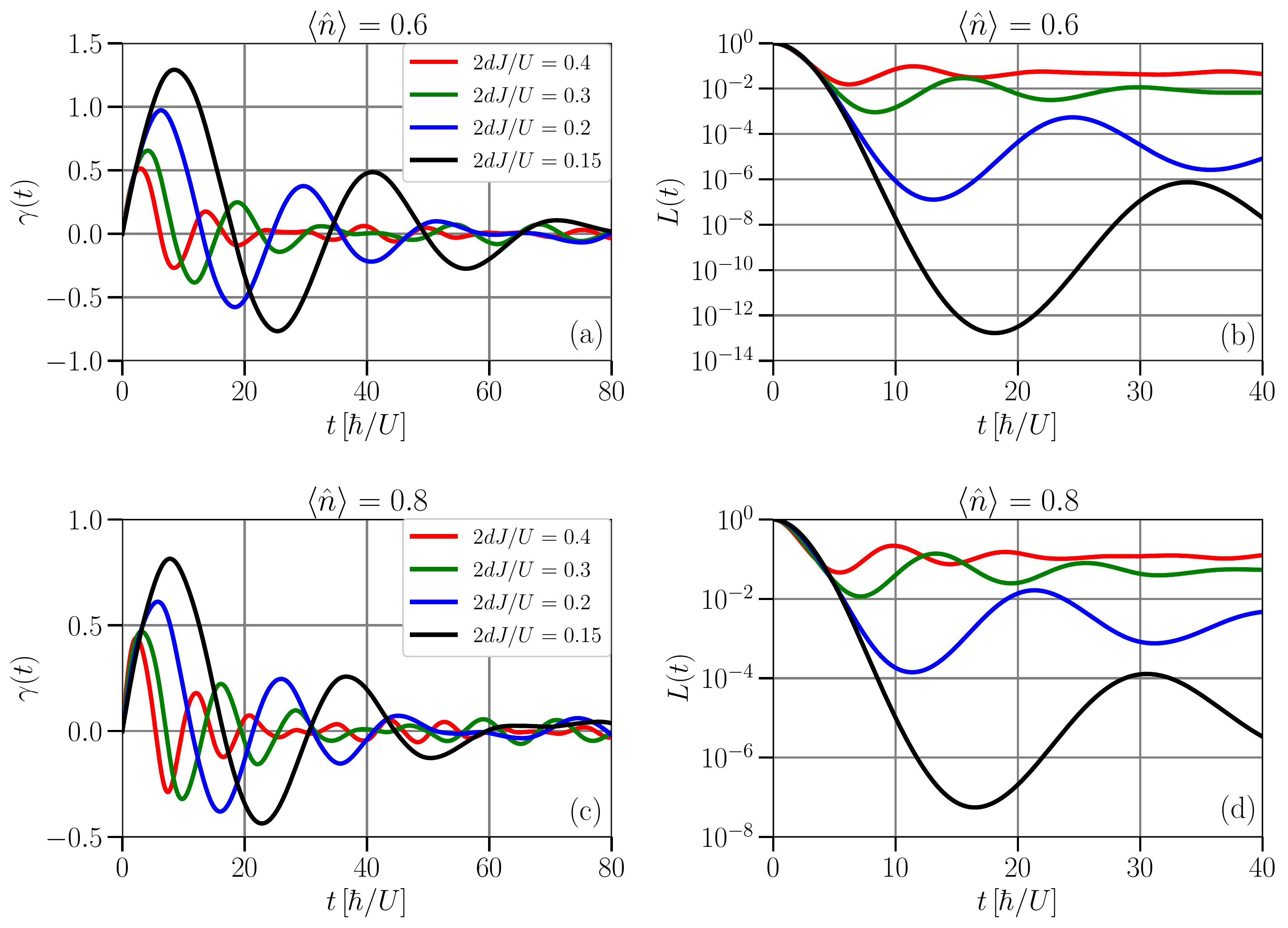}
    \caption{(a) Dephasing rate $\gamma{\left( t \right)}$ given by Eq.~\eqref{gamma_1} at constant density $\langle \hat{n} \rangle = 0.6$ in $d = 2$ dimensions for decreasing $2 \, d \, J/U$ (from red to black solid line) on approaching the hard-core limit of the SF phase.
    (b) Loschmidt echo $L{\left( t \right)}$ corresponding to the the dephasing rates in panel (a).
    (c)-(d) Dephasing rate and Loschmidt echo for the same values of $2 \, d \, J/U$ at a larger, non-integer filling $\langle \hat{n} \rangle = 0.8$.}
    \label{incommensurate_gammas}
\end{figure}

\autoref{incommensurate_gammas}(a) shows the change in the dephasing rate $\gamma{\left( t \right)}$ for decreasing hopping energy $2 \, d \, J/U$ at fixed density $\langle \hat{n} \rangle = 0.6$ (see the corresponding solid black line in \autoref{incommensurate_lines}). We observe that, upon approaching the hard-core limit $2 \, d \, J/U \to 0$ from the deep SF phase, the order of magnitude of $\gamma{\left( t \right)}$ increases significantly, while the time scale of the dephasing dynamics slows down, in such a way that the profiles of $\gamma{\left( t \right)}$ at different values of $2 \, d \, J/U$ are related by a simple scaling relation. On the other hand, the strongly-correlated SF regime still exhibits an evident non-Markovian character, as recognizable also in the oscillating behaviour of the Loschmidt echo $L{\left( t \right)}$, see \autoref{incommensurate_gammas}(b). Here, we can appreciate how non-Markovianity and the overall magnitude of $\gamma{\left( t \right)}$ compete in controlling the amount of dephasing of the impurity. However, at very small $2 \, d \, J/U$, the strong enhancement of the amplitude of $\gamma{\left( t \right)}$ wins over revival effects and induces almost complete dephasing in a small time interval.

These results find an intuitive explanation in the physical properties of the hard-core SF state. For $t \ll 1/J$, strong bath correlations prevent the density excitations induced by the presence of the impurity from leaving a neighbourhood of the impurity itself, therefore leading to the strong-positive density correlations observed in \autoref{incommensurate_gammas}(a). However, being the hard-core phase still coherent in character, hopping process are favoured at larger times and invert the sign of $\gamma{\left( t \right)}$ in analogy with what we observe in the deep SF regime. Therefore, the total amount of dephasing depends on whether local density correlations are sufficiently strong to overcome non-Markovian effects due to long-range coherence.

The dependence of the dephasing rate on the lattice filling can be understood by looking at \autoref{incommensurate_gammas}(c)-(d), referring to a larger filling $\langle \hat{n} \rangle = 0.8$. In particular, we notice that the oscillation amplitude of $\gamma{\left( t \right)}$ and the speed of the dephasing process decreases as the bath density is increased towards the integer value $\langle \hat{n} \rangle = 1$ required for crossing the MI-SF transition.

Finally, we report the remarkable fact that, upon reaching the hard-core SF regime, the Goldstone mode alone still provides the most important part of $\gamma{\left( t \right)}$, which is essentially given by its Gaussian contribution $\gamma_1{\left( t \right)}$ (see the discussion of Section~\ref{app_explicit_gammas}). This implies that a single-particle description of the BH bath is a good approximation for the dephasing dynamics when the impurity is embedded in a strongly-interacting superfluid away from the MI-SF criticality.

\section{Dephasing dynamics in free and weakly-interacting boson baths}\label{app_boson_baths}
The free boson spectrum on the continuum is the Galilean quadratic dispersion relation
\begin{SEquation}\label{e_0}
\varepsilon_0{\left( \mathbf{k} \right)} = \frac{\hbar^2 \, \mathbf{k}^2}{2 \, m} \, ,
\end{SEquation}
so that the spectral function of density correlations scales as $J_0{\left( \omega \right)} \sim \int d^d\mathbf{k} \, \delta{\left[ \omega - \varepsilon_0{\left( \mathbf{k} \right)} \right]} \sim \omega^{(d - 2)/2}$ at small frequencies in $d$ dimensions. It follows that the dephasing rate and decoherence function behave as
\begin{SEquation}\label{gamma_Gamma_0}
\gamma_0{\left( t \right)} = \int d\omega \, \frac{J_0{\left( \omega \right)} \, \sin{\left( \omega \, t \right)}}{\omega} \sim t^{(2 - d)/2} \qquad \Gamma_0{\left( t \right)} = \int d\omega \, \frac{J_0{\left( \omega \right)} \left[ 1 - \cos{\left( \omega \, t \right)} \right]}{\omega^2} \sim t^{(4 - d)/2}
\end{SEquation}
at large times, suggesting that free bosons lead to total dephasing $\exp{\left[ -\Gamma_0{\left( t \to \infty \right)} \right]} = 0$ if $d < 4$. Indeed, the asymptotic behaviour of the dephasing rate $\gamma_0{\left( t \right)} \sim const.$ in $d = 2$ resembles the Markovian behaviour that we observe at the edge transition described in \autoref{SF}(c), where an effective free-particle description of the superfluid phase holds \cite{sachdev, caleffi2019quantum}. A similar result applies to the case of lattice free bosons, for which the spatial discretization introduces only a small, fast-oscillating modulation of $\gamma_0{\left( t \right)}$.

As regards the case of a weakly-interacting gas either on the continuum or on a lattice, within the Bogoliubov approximation the single-particle spectral amplitude of density fluctuations reads $N_{bog, \mathbf{k}} = \sqrt{\rho_0} \left( u_{\mathbf{k}} + v_{\mathbf{k}} \right)$, where $\rho_0$ is the condensate fraction and $u_{\mathbf{k}}$ $\left( v_{\mathbf{k}} \right)$ is the particle (hole) excitation amplitude of the Goldstone mode. Since $N^2_{bog, \mathbf{k}} \sim \left| \mathbf{k} \right|$ at small momenta, we obtain that the low-energy behaviour of the spectral density is controlled by the spatial dimension only,
\begin{SEquation}\label{J_wi}
J_{wi}{\left( \omega \right)} = \int d^d\mathbf{k} N^2_{bog, \mathbf{k}} \delta{\left( \omega - \omega_{bog, \mathbf{k}} \right)} \sim \omega^d \, ,
\end{SEquation}
apart from subdominant corrections depending on the concavity of the Goldstone spectrum $\omega_{bog, \mathbf{k}}$. Equation \eqref{J_wi} leads to
\begin{SEquation}\label{gamma_Gamma_wi}
\gamma_{wi}{\left( t \right)} = \int d\omega \, \frac{J_{wi}{\left( \omega \right)} \, \sin{\left( \omega \, t \right)}}{\omega} \sim t^{-d} \qquad \Gamma_{wi}{\left( t \right)} = \int d\omega \, \frac{J_{wi}{\left( \omega \right)} \left[ 1 - \cos{\left( \omega \, t \right)} \right]}{\omega^2} \sim t^{1 - d}
\end{SEquation}
for large times. Therefore, a weakly-interacting bath induces only partial dephasing, namely $\exp{\left[ -\Gamma_{wi}{\left( t \to \infty \right)} \right]} \neq 0$, at least for $d > 1$.

Most importantly, the frequency dependence of $J_{wi}{\left( \omega \right)}$ on the continuum assures that non-Markovian effects do not occur in any dimension. For instance, for a $d = 1$ gas we find
\begin{SEquation}\label{J_continuum}
J^{1D, cont.}_{wi}{\left( \omega \right)} = \sqrt{\frac{2 \, m}{\hbar^2}} \rho_0 \sqrt{\frac{\sqrt{\left( \rho_0 \, U \right)^2 + \omega^2} - \rho_0 \, U}{\left( \rho_0 \, U \right)^2 + \omega^2}} \sim \sqrt{\frac{m \, \rho_0}{\hbar^2 \, U}} \, \omega \quad \text{for} \ \omega \to 0
\end{SEquation}
which is a monotonous smooth function of $\omega$. On the other hand, for weakly-interacting bosons loaded on a one-dimensional lattice, the spectral density
\begin{SEquation}\label{J_lattice}
\begin{aligned}
J^{1D, latt.}_{wi}{\left( \omega \right)} &= \sqrt{\frac{1}{J}} \rho_0 \sqrt{\frac{\sqrt{\left( \rho_0 \, U \right)^2 + \omega^2} - \rho_0 \, U}{\left( \rho_0 \, U \right)^2 + \omega^2}} \frac{1}{\sqrt{1 - \frac{1}{4 \, J} \left[ \sqrt{ \left( \rho_0 \, U \right)^2 + \omega^2} - \rho_0 \, U \right]}} \\
&\sim \sqrt{\frac{\rho_0}{2 \, J \, U}} \omega \quad \text{for} \ \omega \to 0
\end{aligned}
\end{SEquation}
presents a van Hove singularity where the dispersion relation of the Goldstone mode reaches a stationary point, namely at the boundary of the Brillouin zone $k = \pi$, where $\omega = \sqrt{2 \, J \left( 2 \, J + 2 \, \rho_0 \, U \right)}$. This change in the high-energy structure of $J_{wi}{\left( \omega \right)}$ is a genuine effect of the absence of full Galilean invariance due to spatial discreteness inherent to the lattice: in fact, the lattice setting introduces an additional energy scale fixed by the bandwidth of the Goldstone excitation, approximately proportional to the hopping energy $J$ in the weakly-interacting limit $J/U \gg 1$. Consequently, passing from the continuum to the lattice, in the superfluid phase the dephasing function $\gamma{\left( t \right)}$ acquires an oscillating behaviour whose period is set by the hopping time scale, as we observe e.g. in the 2D result shown in \autoref{SF}(a). On the other hand, the amplitude of the oscillations of $\gamma{\left( t \right)}$ at large times is always controlled by the power-law decay \eqref{gamma_Gamma_wi} seen on the continuum.

\bigskip

\begin{figure}[!htp]
    \centering
	\includegraphics[width=0.92\linewidth]{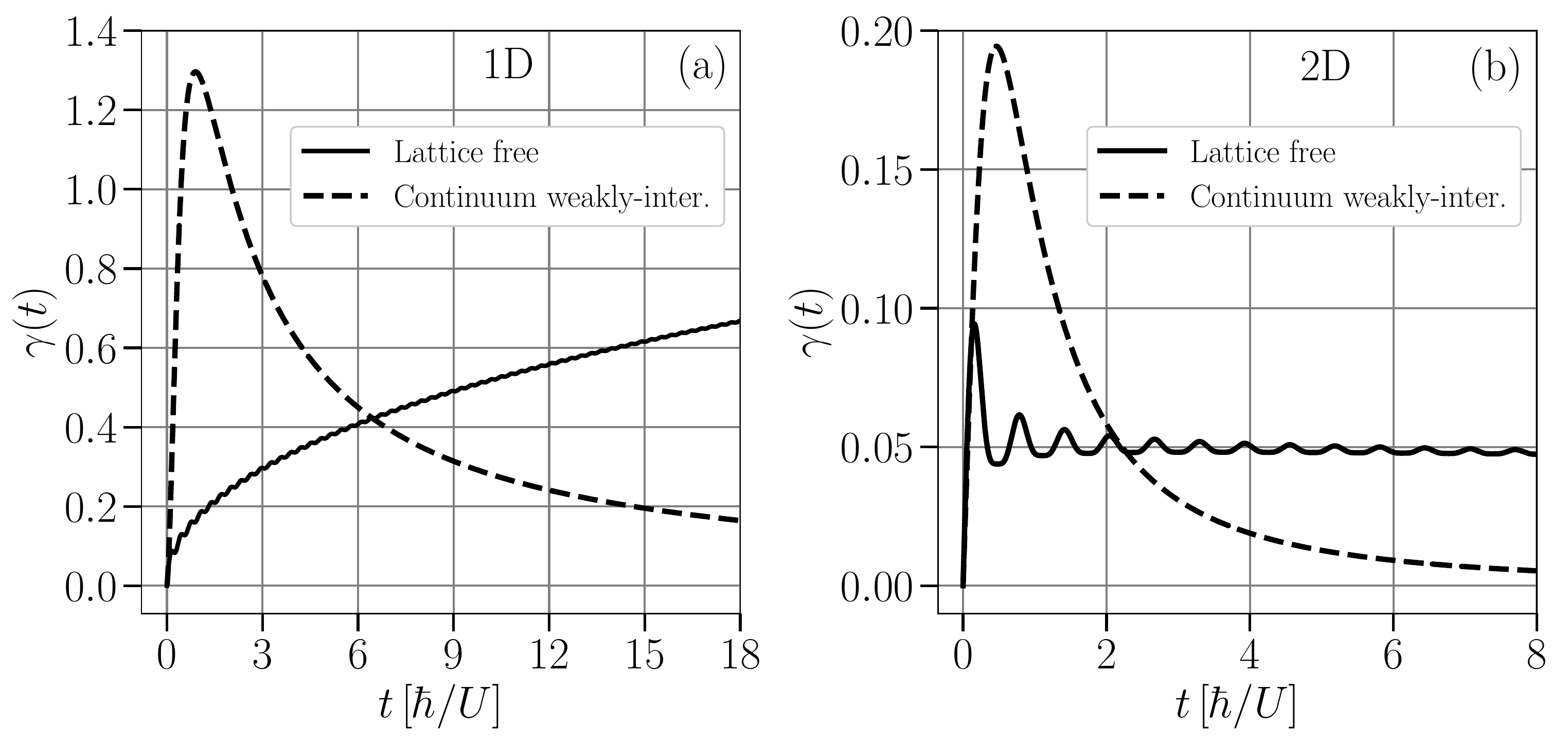}
    \caption{(a) Dephasing rate $\gamma{\left( t \right)}$ for a 1D bath of lattice free bosons (solid line) and weakly-interacting bosons on the continuum (dashed line). Notice the $\sqrt{t}$ growth for free particles and the $t^{-1}$ decay for weak interactions.
    (b) The same quantities calculated for $d = 2$. Notice the constant-value asymptotics of $\gamma{\left( t \right)}$ in presence of free bosons and the $t^{-2}$ decay for weak interactions.}
    \label{fig_SM_D}
\end{figure}

\autoref{tab1} summarises the previous discussion and reports the expressions of $J{\left( \omega \right)}$ and $\gamma{\left( t \right)}$ for the most relevant cases and limits. For the sake of completeness, \autoref{fig_SM_D} reports the behaviour of the dephasing rate $\gamma{\left( t \right)}$ for a bath of free (weakly-interacting) bosons loaded on a square lattice (on the continuum), to be compared with our results for the critical SF phase of the BH bath at the edge transition. Finally, \autoref{tab2} displays the long-time behaviour of the decoherence function $\Gamma{\left( t \right)}$ and of the Loschmidt echo $L{\left( t \right)}$ for the same reference cases.

\begin{table}
\footnotesize
\begin{center}
\begin{tabular}{|m{1.72cm}||c|c|}
\hline
& $\mathbf{J{\left( \omega \right)}}$ & $\mathbf{\gamma{\left( t \right)} = d\Gamma{\left( t \right)}/dt}$ \\
\hline
Continuum free bosons & $\omega^{\left( d - 2 \right)/2}$ & $t^{\left( 2 - d \right)/2} \quad \text{for} \quad 0 < d < 4$ \\
\hline
Lattice free bosons (1D) & $\left[ \frac{\omega}{J} \left( 1 - \frac{\omega}{4 \, J} \right) \right]^{-1/2}$ & $t \left[ \sin{\left( 2 \, J \, t \right)} \, J_1{\left( 2 \, J \, t \right)} + \cos{\left( 2 \, J \, t \right)} \, J_0{\left( 2 \, J \, t \right)} \right] \hspace{0.2cm} *$ \\
\hline
Lattice free bosons & $\omega^{\left( d - 2 \right)/2} \quad \text{for} \quad \omega \ll J$ & $t^{\left( 2 - d \right)/2} \quad \text{for} \quad 0 < d < 4 \hspace{0.2cm} *$ \\
\hline
Continuum weakly interacting bosons & $\frac{\left[ \sqrt{\left( \rho_0 \, U \right)^2 + \omega^2} - \rho_0 \, U \right]^{d/2}}{2 \sqrt{\left( \rho_0 \, U \right)^2 + \omega^2}}$ & $t^{-d} \quad \text{for} \quad t \gg m/\hbar^2$ \\
\hline
Lattice weakly interacting bosons (1D) & $\sqrt{\frac{\sqrt{\left( \rho_0 \, U \right)^2 + \omega^2} - \rho_0 \, U}{\left( \rho_0 \, U \right)^2 + \omega^2}} \frac{1}{\sqrt{1 - \frac{1}{4 \, J} \left[ \sqrt{ \left( \rho_0 \, U \right)^2 + \omega^2} - \rho_0 \, U \right]}}$ & $t^{-1} \quad \text{for} \quad t \gg 1/J$ \\
\hline
Lattice weakly interacting bosons & $\frac{\left[ \sqrt{\left( \rho_0 \, U \right)^2 + \omega^2} - \rho_0 \, U \right]^{d/2}}{\sqrt{\left( \rho_0 \, U \right)^2 + \omega^2}} \quad \text{for} \quad \omega \ll J$ & $t^{-d} \quad \text{for} \quad t \gg 1/J$ \\
\hline
\end{tabular}
\end{center}
\caption{$\star$ In the case of free bosons loaded on a 1D lattice, the dephasing rate behaves as $\gamma{\left( t \right)} = t \left[ \sin{\left( 2 \, J \, t \right)} \, J_1{\left( 2 \, J \, t \right)} + \cos{\left( 2 \, J \, t \right)} \, J_0{\left( 2 \, J \, t \right)} \right] \sim \sqrt{t}$ on a coarse-grained time scale (i.e. for $t \gg 1/J$), modulated by small oscillations due to the lattice discretization [see \autoref{fig_SM_D}(a)]. Therefore, in the long-time limit, the pure dephasing dynamics in a free-boson environment is insensitive to the spatial discretization due to the lattice. We report the same dynamical behaviour for $d > 1$.}
\label{tab1}
\end{table}

\begin{table}
\footnotesize
\begin{center}
\begin{tabular}{|m{1.72cm}||c|c|} 
\hline
& $\mathbf{\Gamma{\left( t \right)}}$ & $\mathbf{L{\left( t \right)}}$ \\
\hline
Continuum free bosons & $t^{\left( 4 - d \right)/2} \quad \text{for} \quad 0 < d < 4$ & $\exp{\left[ -\beta \, t^{\left( 4 - d \right)/2} \right]} \quad \text{for} \quad 0 < d < 4$ \\
\hline
Lattice free bosons (1D) & $t^{3/2} \hspace{0.2cm}$ & $\exp{\left( -\beta \, t^{3/2} \right)} \hspace{0.2cm}$ \\
\hline
Lattice free bosons & $t^{\left( 4 - d \right)/2} \quad \text{for} \quad 0 < d < 4 \quad$ & $\exp{\left[ -\beta \, t^{\left( 4 - d \right)/2} \right]} \quad \text{for} \quad 0 < d < 4 \quad$ \\
\hline
Continuum weakly interacting bosons & $\begin{aligned} &\ln{\left( t \right)} \quad \text{for} \quad d = 1 \\ &\, t^{1 - d} \quad \text{for} \quad d > 1 \end{aligned} \quad \text{for} \quad t \gg m/\hbar^2$ & $\begin{aligned} &t^{-\alpha} \quad \text{with} \quad \alpha > 0 \quad \text{for} \quad d = 1 \\ &\exp{\left( -\beta \, t^{1 - d} \right)} \quad \text{for} \quad d > 1 \end{aligned} \quad \text{for} \quad t \gg m/\hbar^2$ \\
\hline
Lattice weakly interacting bosons (1D) & $\ln{\left( t \right)} \quad \text{for} \quad t \gg 1/J$ & $t^{-\alpha} \quad \text{with} \quad \alpha > 0 \quad \text{for} \quad t \gg 1/J$ \\
\hline
Lattice weakly interacting bosons & $t^{1 - d} \quad \text{for} \quad t \gg 1/J$ & $\exp{\left[ -\beta \, t^{1 - d} \right]} \quad \text{for} \quad t \gg 1/J$ \\
\hline
\end{tabular}
\end{center}
\caption{}
\label{tab2}
\end{table}

\normalsize

\pagebreak
\bibliographystyle{apsrev4-1}
\bibliography{ref}

\end{document}